\documentclass[twocolumn,tightenlines,showpacs,prd,floatfix,preprintnumbers,amsmath,amssymb,nofootinbib,superscriptaddress]{revtex4}
\usepackage{graphicx}
\usepackage{subfigure}

\def\pheq{\phantom{=}}
\newcommand{\Deqn}[1]{{Eq.~(\ref{#1})}}
\newcommand{\Deqns}[1]{{Eqs.~(\ref{#1})}}

\newcommand{\beq}{\begin{equation}}
\newcommand{\eeq}{\end{equation}}
\newcommand{\bea}{\begin{eqnarray}}
\newcommand{\eea}{\end{eqnarray}}
\newcommand{\tr}{{\mathop{\mathrm{tr}}\nolimits}}

\begin{document}

\title{Optimal strategies for 
gravitational wave stochastic background searches in pulsar timing data}

\author{Melissa Anholm} 
\email{anholm@gravity.phys.uwm.edu}
\affiliation{Center for Gravitation and Cosmology, Department of Physics, 
University of Wisconsin--Milwaukee, P.O. Box 413, Milwaukee, Wisconsin 53201, 
USA}
\author{Stefan Ballmer}
\email{sballmer@caltech.edu}
\affiliation{LIGO Laboratory, California Institute of Technology, MS 18-34,
Pasadena, California 91125, USA}
\author{Jolien D. E. Creighton}
\email{jolien@gravity.phys.uwm.edu}
\affiliation{Center for Gravitation and Cosmology, Department of Physics, 
University of Wisconsin--Milwaukee, P.O. Box 413, Milwaukee, Wisconsin 53201, 
USA}
\author{Larry R. Price} 
\email{larry@gravity.phys.uwm.edu}
\affiliation{Center for Gravitation and Cosmology, Department of Physics, 
University of Wisconsin--Milwaukee, P.O. Box 413, Milwaukee, Wisconsin 53201, 
USA}
\author{Xavier Siemens}
\email{siemens@gravity.phys.uwm.edu}
\affiliation{Center for Gravitation and Cosmology, Department of Physics, 
University of Wisconsin--Milwaukee, P.O. Box 413, Milwaukee, Wisconsin 53201, 
USA}

\date{\today}

\begin{abstract} 

  A low frequency stochastic background of gravitational waves may be
  detected by pulsar timing experiments in the next five to ten years.
  Using methods developed to analyze interferometric gravitational
  wave data, in this paper we lay out the optimal techniques to detect
  a background of gravitational waves using a pulsar timing array. We
  show that for pulsar distances and gravitational wave frequencies
  typical of pulsar timing experiments, neglecting the effect of the
  metric perturbation at the pulsar does not result in a significant
  deviation from optimality. We discuss methods for setting upper
  limits using the optimal statistic, show how to construct skymaps
  using the pulsar timing array, and consider several issues associated
  with realistic analysis of pulsar timing data.

\end{abstract}

\maketitle

\section{Introduction}

The search for gravitational waves is at the forefront of current
fundamental physics research. The direct detection of gravitational
waves will usher in a new era in astronomy and astrophysics.
Gravitational waves will reveal information about black holes,
supernovae, and neutron stars that cannot be gleaned from
electromagnetic observations. Furthermore, the detection of a
gravitational wave background will open an observational window onto a
time in the early universe before recombination, prior to which the
universe is opaque to electromagnetic waves. The scientific rewards
for such a detection would be truly exceptional.  Several
international efforts are underway to detect gravitational waves and
two of these efforts are expected to result in the detection of
gravitational waves in the next 5 to 10 years: Interferometric
ground-based gravitational wave detectors and pulsar timing
observations.

Neutron stars radiate powerful beams of radio waves from their
magnetic poles. If a neutron star's magnetic poles are not aligned
with its rotational axis, the beams sweep through space like the
beacon on a lighthouse and the neutron star is said to be a pulsar. If
the Earth lies within the sweep of a pulsar's beams, the star is
observed as a point source in space emitting short, rapid bursts of
radio waves~\cite{Hulse:1974eb}.  Due to their enormous mass, neutron
stars have a very large moment of inertia and the radio pulses we
observe arrive at a very constant rate.  Pulsar timing experiments
exploit this
regularity~\cite{1975GReGr...6..439E,Sazhin78,Detweiler,%
LorimerKramer2005}. Fluctuations in the time of arrival of radio
pulses, after all known effects have been subtracted out, could be due
to the presence of gravitational waves. Since the 1970s, when these
ideas were first conceived, pulsar timing precision has improved
dramatically. Several known pulsars can now be timed with a precision
of about 1 micro-second, and a handful can be timed with a precision
around several hundred nanoseconds \cite{Hobbs:2005yx}. Recent
work~\cite{Jenet:2005pv} has shown that the presence of nanohertz
gravitational waves could be detected by observing 20 pulsars with
timing precisions of 100 nanoseconds over a period of 5 to 10 years.
Non-detection would still improve current bounds on the low frequency
stochastic gravitational wave background~\cite{Jenet:2006sv}.

Gravitational waves from supermassive black
hole binary systems could be detected via pulsar timing
observations~\cite{Wyithe:2002ep,Jaffe:2002rt,Enoki:2004ew,%
Kramer:2004hd,alberto}. In addition, pulsar timing has the potential to
measure the polarization properties of gravitational waves which could
confirm (or even change) the current theory of
gravity~\cite{PhysRevLett.30.884,PhysRevD.8.3308}. Gravitational wave 
observations in the nanohertz band could also
yield information about the early universe~\cite{Boyle:2007zx}. Cosmic
strings, line-like topological defects, could produce gravitational
waves in the nanohertz band. Cosmic strings can form during phase
transitions in the early universe due to the rapid cooling that takes
place after the Big Bang~\cite{Kibble:1976sj,Hindmarsh:1994re,%
  alexbook}. Cosmic string production is generic in supersymmetric
grand unified theories~\cite{Jeannerot:2003qv}. Additionally, in
string theory motivated cosmological models cosmic strings may also
form (dubbed cosmic superstrings to differentiate them from
field theoretic cosmic strings)~\cite{Jones:2002cv,%
Sarangi:2002yt,Dvali:2003zj,Jones:2003da,Copeland:2003bj,%
Jackson:2004zg}. Cosmic strings and superstrings are expected to
produce a background of stochastic gravitational waves and bursts of
gravitational waves~\cite{Berezinsky:2000vn,%
Damour:2000wa,Damour:2001bk,Damour:2004kw,Siemens:2006vk,%
Siemens:2006yp} that could be detected using pulsar timing
observations. Pulsar timing observations are already producing some of
the most interesting constraints on cosmic string models and a
detection would have profound implications~\cite{Siemens:2006vk,%
Siemens:2006yp}.

Recently Lommen~\cite{Lommen:2002je} produced an upper limit on the
stochastic gravitational wave background using observations of three
millisecond pulsars spanning 17 years. The methods used by Lommen were
based on those developed by Kaspi and
collaborators~\cite{Kaspi:1994hp} and have been the subject of some
criticism in the literature~\cite{Damour:2004kw,Jenet:2006sv}.  More
recently Jenet and collaborators~\cite{Jenet:2005pv,Jenet:2006sv}
developed a new technique for gravitational wave stochastic background
searches in pulsar timing data and applied it to Parkes Pulsar Timing
Array data~\cite{Manchester:2007mx,PPTA}.

In this paper we consider optimal strategies for extraction of
a gravitational wave stochastic background signal using 
data from a pulsar timing array. Our methods 
are based on those developed for and used in ground based 
interferometric gravitational-wave detectors such as LIGO and
Virgo~\cite{Michelson,Christensen:1992wi,Flanagan:1993ix,%
AllenandRomano,Abbott:2003hr,%
Abbott:2005ez,Abbott:2006zx,Abbott:2007tw,Abbott:2007wd}, and improve
on existing methods in several ways. In Section~\ref{sigsec} we write the
redshift of pulsar signals induced by passing gravitational waves, first
derived by Detweiler~\cite{Detweiler}, in a coordinate-independent way
more suitable for our analysis, and discuss its form in the frequency
domain including the long-wavelength limit.  In Section~\ref{DSsec} we
construct the optimal cross-correlation filter for a pulsar pair by
maximizing the signal to noise. We find that the overlap reduction
function is well approximated by a constant, or equivalently, that the
metric perturbation at the pulsar can be neglected for values of
pulsar distances and gravitational wave frequencies typical of pulsar
timing experiments, without significant losses in sensitivity.  
In Section~\ref{DSsec} we also show how to construct the
optimal combination of cross-correlations of pulsar pairs in a pulsar
timing array and include a more sophisticated derivation of the
optimal detection statistic based on the likelihood ratio. In Section~\ref{ULDsec}
we discuss upper limit and detection
methods. In Section~\ref{SMsec} we show how to construct skymaps using pulsar
timing data---a pulsar timing radiometer. In Section~\ref{issues} we discuss
several important issues relating to the realistic analysis of pulsar
timing data, including the Lomb-Scargle periodogram for power spectrum
estimation of unevenly sampled data, and optimal procedures for
computing Fourier transforms. We conclude in Section VIII. Lommen,
Romano and Woan~\cite{Lommenea} will extend our work using a
likelihood based approach developed in~\cite{Allen:2002jw}, and
consider the case of stochastic backgrounds that are loud compared to
the noise, closely examine time-domain implementations of the optimal
statistic, and provide a detailed comparison of the optimal statistic
described here with the methods of Jenet and
collaborators~\cite{Jenet:2005pv,Jenet:2006sv}.

\section{\label{sigsec}The signal}

Gravitational waves affect pulsar timing measurements by creating 
perturbations in the null geodesics that the radio signals emitted from the
pulsar travel on~\cite{Detweiler}.  In this section we will  describe 
the relationship between the metric perturbation and the signal measured in
pulsar timing experiments.

A metric perturbation in a spatial, transverse, traceless gauge has a plane
wave expansion given by~\cite{AllenandRomano}
\beq
h_{ij}(t,\vec{x}) = \sum_{A}
\int_{-\infty}^{\infty}df\, \int_{S^2}d\hat{\Omega}\, 
e^{i2\pi f(t-\hat{\Omega}\cdot\vec{x})}
h_A(f,\hat{\Omega})e_{ij}^A(\hat{\Omega}),
\label{pwexp}
\eeq
where $f$ is the frequency of the gravitational waves, 
$\vec k= 2 \pi f \hat \Omega$ is the wave vector, $\hat \Omega$ is a unit vector
that points along the direction of travel of the waves, $i,j=x,y,z$
are spatial indices, and
the index $A=+,\times$ labels polarizations.
The polarization tensors $e_{ij}^A(\hat{\Omega})$ are
\begin{subequations}
\begin{align}
\label{eq:e_plus}
  e_{ij}^+({\hat{\Omega}}) &=  {\hat{m}}_i {\hat{m}}_j - {\hat{n}}_i {\hat{n}}_j,\\
\label{eq:e_cross}
  e_{ij}^{\times}({\hat{\Omega}}) &= {\hat{m}}_i {\hat{n}}_j + {\hat{n}}_i {\hat{m}}_j,
\end{align}
\end{subequations}
where
\begin{subequations}
\begin{align}
\label{eq:omega}
  {\hat{\Omega}}&= (\sin{\theta} \cos{\phi},  \sin{\theta} \sin{\phi},  
\cos{\theta}), \\
\label{eq:m}
  {\hat{m}}&=(\sin{\phi}, -\cos{\phi}, 0), \\
\label{eq:n}
  {\hat{n}}&=(\cos{\theta}\cos{\phi}, \cos{\theta}\sin{\phi}, -\sin{\theta}).
\end{align}
\end{subequations}

Now consider the metric perturbation from a single gravitational wave 
traveling along the $z$-axis so that $\hat{\Omega}=\hat{z}$.  The metric 
perturbation is given explicitly by
\bea
h_{ij}(t,\hat{\Omega}=\hat{z}) &=& \sum_{A}
\int_{-\infty}^{\infty}df\,
e^{i2\pi f(t-z)}
h_A(f,\hat{z})e_{ij}^A(\hat{z})
\nonumber\\
&\equiv& h_{ij}(t-z).
\label{mpzdir}
\eea
The physical metric due to the perturbation is given by
\begin{equation}
\label{gmunu}
g_{ab}=\eta_{ab}+h_{ab}(t-z)=
\left(
\begin{array}{cccc}
-1 & 0 & 0 & 0  \\
0 & 1+h_+ & h_{\times} & 0 \\
0 & h_{\times} & 1-h_+ & 0 \\
0 & 0 & 0 & 1\\
\end{array}
\right),
\end{equation}
where $\eta_{ab}={\rm diag}\{-1,1,1,1\}$ is the Minkowski metric,
$a,b$ are spacetime indices, and
\beq
h_{+,\times} =h_{+,\times}(t-z)= \int_{-\infty}^{\infty}df\, e^{i2\pi f(t-z)}
h_{+,\times}(f,\hat{z}).
\label{hpcdef}
\eeq
In this background, a pulsar emitting pulses at frequency $\nu_0$ and 
direction cosines $\alpha$, $\beta$, and $\gamma$, with respect to 
the $x$-, $y$-, and $z$-axes, respectively, will be observed to 
change its frequency in the solar system reference frame
according to~\cite{Detweiler}
\begin{eqnarray} 
\label{redshift}
z(t,\hat z) &\equiv& \frac{\nu_0-\nu(t)}{\nu_0}
\nonumber
\\
&=& \frac{\alpha^2-\beta^2}{2(1+\gamma)}(h_{+}^{\rm p}-h_{+}^{\rm e})
+ \frac{\alpha\beta}{1+\gamma} (h_{\times}^{\rm p}-h_{\times}^{\rm e}),
\end{eqnarray} 
where $h_{+,\times}^{\rm e}$, $h_{+,\times}^{\rm p}$ are the gravitational wave
strains at the solar-system barycenter and the pulsar, respectively. 
This central result was obtained by Detweiler~\cite{Detweiler}, 
who generalized a result of Estabrook and 
Wahlquist~\cite{1975GReGr...6..439E} to include both gravitational 
wave polarizations and for pulsars at
arbitrary locations. They in turn based their calculation on an
earlier one by Kaufmann~\cite{Kaufmann}.  A detailed derivation of this result is 
provided in Appendix \ref{Detderiv}.

Looking at Eq.~(\ref{redshift}) (and as shown in Appendix~\ref{app:lin}) we can write the 
redshift $z(t,\hat \Omega)$ of signals from a pulsar in the  
direction of the unit vector $\hat p$ produced by a gravitational 
wave coming from the direction $\hat \Omega$ as, 
\beq
z(t,\hat{\Omega}) = 
\frac{1}{2}
\frac{\hat{p}^i\hat{p}^j}{1+\hat{\Omega}\cdot\hat{p}}\Delta h_{ij},
\label{eqzsom}
\eeq
where
\beq
\Delta h_{ij}
\equiv
h_{ij}(t_{\rm p},\hat{\Omega}) - 
h_{ij}(t_{\rm e},\hat{\Omega}),
\label{delhdef}
\eeq
is the difference in the metric perturbation traveling along the direction 
$\hat \Omega$ at the pulsar and at the center of the solar system.
The vectors $(t_{\rm e}, \vec{x}_{\rm e})$ and $(t_{\rm p},\vec{x}_{\rm p})$ 
give the spacetime coordinates of the solar-system barycenter 
and the pulsar, respectively. The metric perturbation at each 
location takes the form,
\beq
h_{ij}(t,\hat{\Omega}) = \sum_{A}
\int_{-\infty}^{\infty}df\, e^{i2\pi f(t-\hat{\Omega}\cdot\vec{x}_0)}
h_A(f,\hat{\Omega})e_{ij}^A(\hat{\Omega}),
\label{mpform}
\eeq
for a fixed $\vec{x}_0$.

We choose a particular coordinate system by placing the solar-system
barycenter at the origin and the pulsar some distance $L$ away.  With
these conventions
\bea
t_{\rm p} &=& t_{\rm e} -L\equiv t-L, \\
\vec{x}_{\rm e} &=& 0, \\
\vec{x}_{\rm p} &=& L\hat{p}.
\eea
If assume that the amplitude of the metric perturbation is the same at the 
solar-system
barycenter and the pulsar then we can use Eq.~(\ref{mpform}) to write out 
$\Delta h_{ij}$ in our coordinate system as
\bea
\Delta h_{ij} 
&=& \int_{-\infty}^{\infty}df\, 
e^{i2\pi f t}\left(
e^{-i2\pi f L\left(1+\hat{\Omega}\cdot\hat{p}\right)}-1\right)
\nonumber\\ 
&\pheq& \phantom{\int_{-\infty}^{\infty}df\,}
\times \sum_A h_A(f,\hat{\Omega})e_{ij}^A(\hat{\Omega})
\nonumber\\
&\equiv&\Delta h_{ij}(t,\hat{\Omega}).
\eea
Ultimately, we will be interested in the Fourier transform of this quantity
which is simply
\beq
\Delta \tilde{h}_{ij}(f,\hat{\Omega}) 
= \left(
e^{-i2\pi f L\left(1+\hat{\Omega}\cdot\hat{p}\right)}-1\right)
\sum_A h_A(f,\hat{\Omega})e^A_{ij}(\hat{\Omega}).
\eeq
We can then write the Fourier 
transform of Eq.~(\ref{eqzsom}) as
\beq
\tilde{z}(f,\hat{\Omega}) 
= \left(e^{-i2\pi f L(1+\hat{\Omega}\cdot\hat{p})}-1\right)
\sum_{A} h_A(f,\hat{\Omega}) F^A(\hat{\Omega}),
\label{zredshift1}
\eeq
where we have defined
\beq
F^A(\hat{\Omega})\equiv e^A_{ij}(\hat{\Omega}) \frac{1}{2}
\frac{\hat{p}^i\hat{p}^j}{1+\hat{\Omega}\cdot\hat{p}}.
\label{eqFA}
\eeq
As shown in Appendix~\ref{app:lin} 
the total redshift is given by summing over the contributions coming from 
gravitational waves in every direction:
\bea
\tilde{z}(f) = \int_{S^2} d\hat{\Omega}\, \tilde{z}(f,\hat{\Omega}),
\label{zftot}
\eea
and similarly for $z(t)$. 

In fact, it is not the redshift, but a related quantity called the 
\emph{residual} that gets reported in pulsar timing measurements.  The 
residual, $R(t)$, is defined as the integral of the redshift:
\beq
R(t)\equiv \int_0^t dt'\, z(t').
\eeq
This simple relationship gives us the freedom to develop the data analysis
for either variable and we henceforth limit our attention to the redshift,
but the results here can be phrased in terms of the residual with minimal 
effort.

In the literature, searches for gravitational waves using pulsar
timing data are typically performed in the time domain. The (unknown)
metric perturbation at the pulsar in, say, Eqs.~(\ref{redshift}) or
(\ref{eqzsom}) is neglected because one can treat it as another noise
term which averages to zero when performing correlations between
measurements of different pulsars. In the frequency domain this is unnecessary.
Eq.~(\ref{zredshift1}) does not depend explicitly on the metric
perturbation at the pulsar, rather the dependence is all in a distance 
and frequency dependent phase factor.  It is then conceivable that if we
could determine the distance to a pulsar $L$ with sufficient accuracy
we could use the metric perturbation at the pulsar to improve the
sensitivity of our searches.  Unfortunately, such measurements of
pulsar distances are unavailable. We will show in Section~\ref{OFsec},
however, that for the case of a stochastic background search in pulsar
timing data the phase factor can be neglected without any significant
loss in sensitivity.  It is unclear whether this is true for
other types of gravitational wave searches.

From the ground-based interferometer perspective Eqs.~(\ref{redshift}) or
(\ref{eqzsom}) are somewhat counter-intuitive. This difficulty arises 
from the factor of $1+\hat \Omega \cdot \hat p$ in the denominator; in 
Appendices~\ref{Detderiv} and \ref{app:lin} we show explicitly how
this factor enters 
the expression. When $\hat \Omega \cdot
\hat p = \pm 1$, i.e. the gravitational wave and the pulsar directions
are parallel or anti-parallel, Eqs.~(\ref{zredshift1}) and (\ref{eqFA})
lead to no redshifting of the pulsar signal for completely different
reasons.  When they are parallel the reason is the transverse nature
of gravitational waves, and when they are anti-parallel it is because
the pulsar signals ``surf'' the gravitational waves. Our surprise is a
result of our long-wavelength limit intuition.

Eq.~(\ref{zredshift1}) has an obvious long-wavelength limit. We can
use this limit to compare the form of our results with those of
ground-based interferometers such as LIGO. When $2 \pi f L \ll 1$ we
can Taylor expand the exponential and to first order
Eq.~(\ref{zredshift1}) becomes 
\beq 
\tilde{z}(f,\hat{\Omega})
 \approx -i\pi f L \hat{p}^i\hat{p}^j
\sum_{A} h_A(f,\hat{\Omega}) e^A_{ij}(\hat{\Omega}).
\label{zredshift2}
\eeq
Typical values of $f$ are in the range $1/10$~yr$^{-1}$ to
$10$~yr$^{-1}$. Typical values of the Earth pulsar distance $L$ are in
the range $100$~ly to $10^4$~ly. This means $f L$ is in the range $10$
to $10^5$ and pulsar timing experiments are never in the long
wavelength limit. However, the Taylor expansion can also be done for
large $f L$ when the angle between $\hat \Omega$ and $\hat p$ is
sufficiently close to $\pi$. In this case the pulsar signals can ``surf'' the
gravitational waves and not undergo redshifting.  Writing that angle as $\pi
-\epsilon$ with $\epsilon \ll 1$, then the Taylor expansion is also
valid when $\epsilon \ll (\pi f L)^{-1/2}$.

Taking the inverse Fourier transform of Eq.~(\ref{zredshift2}) yields
\beq
\tilde{z}(t) 
 \approx -\frac{L}{2} \hat{p}^i\hat{p}^j \dot h_{ij}(t,\vec x_{\rm e}),
\label{zredshift3}
\eeq
which is the projection of the time derivative of the metric
perturbation at the solar-system barycenter onto the unit vector that
points to the pulsar.  Note that unlike Eq.~(\ref{eqzsom}), this
equation no longer depends on the direction of the gravitational wave and can
be expressed in terms of the full metric perturbation (derivative).
For the case of ground based interferometers the signal, the so-called
strain, is proportional to the difference in length of the two arms
because the signal at the dark port of the interferometer depends on
that difference.  If the arms point in the directions of the unit
vectors $\hat X$ and $\hat Y$ the strain is given by
\beq
h(t)\equiv h_{ij}(t,\vec{x}) \frac{1}{2}(\hat{X}^i\hat{X}^j-\hat{Y}^i\hat{Y}^j),
\eeq
which is the metric perturbation $h_{ij}(t,\vec{x})$ projected onto the 
difference of the arms. 

\section{\label{DSsec}Detection statistic}

With an understanding of the signal in hand we now turn our attention
to developing an optimal detection strategy.  In this section we will
first derive the optimal cross-correlation statistic for a single
pulsar pair using arguments based on maximizing signal to noise ratio.
We will then determine the best way to combine measurements from
multiple pulsar pairs to obtain the most constraining upper limit.
This section will conclude with a more sophisticated derivation of the
optimal detection statistic based on the likelihood ratio.

\subsection{\label{OFsec}The optimal filter}
In this section we will derive the optimal filter for detecting a stochastic
background of gravitational waves from the cross-correlation of
redshift measurements of two 
different pulsars.  This problem was addressed in detail by Allen and 
Romano~\cite{AllenandRomano}, for the case of interferometric
gravitational wave detectors 
and our analysis follows theirs closely.

Consider the signals from two pulsars
\bea
s_1(t)&=&z_1(t)+n_1(t),\\
s_2(t)&=&z_2(t)+n_2(t),
\eea
where $z_i(t)$ is the redshift and $n_i(t)$ is the noise intrinsic in the
measurement.  Throughout this work we will assume that each $n_i(t)$ is
stationary and Gaussian, and is greater in magnitude than the redshift.
Additionally we assume that 
\bea
\langle n_i(t)\rangle &=& 0,
\nonumber
\\
 \langle z_i(t)\rangle&=&0,
\nonumber
\\
\langle n_1(t)n_2(t)\rangle &=&0,
\nonumber
\\
\langle n_i(t)z_j(t)\rangle &=&0,
\label{noiseeqs}
\eea
for all $i$ and $j$, where the angle brackets denote an expectation value.

 A stochastic background will show up in the data as 
correlated noise between measurements with different detectors.  Our goal is to
find a filter, $Q(t-t')$, that optimizes the cross-correlation statistic
\beq
S\equiv \int_{-T/2}^{T/2}dt\,\int_{-T/2}^{T/2}dt'\, s_1(t)s_2(t')Q(t-t'),
\label{eqStd}
\eeq
where $T$ is the observation time.  We will define the
optimal filter to be the $Q(t-t')$ that maximizes the signal to noise ratio (SNR)
\beq
{\rm SNR}\equiv \frac{\mu}{\sigma},\label{snrdef}
\eeq
where $\mu$ and $\sigma$ are the mean and square root of the 
variance, respectively, associated with the cross-correlation 
signal defined in \Deqn{eqStd}.  

We start by assuming that the observation time is much greater than
the separation of the two detectors and extend the limits of the integral over
$dt'$ to $\pm\infty$. Technically our assumption is not correct
because pulsars are typically separated by distances far greater than the
observation time.  Later we will see that neglecting the
phase terms that correspond to the metric perturbation at the pulsar
location is an excellent approximation.  In effect this makes our 
detectors co-located though not co-aligned, and our assumption 
about the observation time is appropriate. We work in the frequency domain
so that~\Deqn{eqStd} becomes
\beq
S=\int_{-\infty}^{\infty}df\, \int_{-\infty}^{\infty}df'\, \delta_T(f-f')
\tilde{s}_1^*(f)\tilde{s}_2(f')\tilde{Q}(f'),
\eeq
where $\delta_T(f-f')$ is the finite-time approximation to the delta function
\beq
\delta_T(f)=\frac{\sin(\pi fT)}{\pi f}.
\eeq

The mean of the cross-correlation is
\beq
\mu\equiv\langle S\rangle = \int_{-\infty}^{\infty}df \int_{-\infty}^{\infty}
df'\, \delta_T(f-f')\langle\tilde{z}_1^*(f)\tilde{z}_2(f')\rangle\tilde{Q}(f').
\label{eqmu}
\eeq
Because of \Deqns{zredshift1}, (\ref{zftot}), and~(\ref{noiseeqs}), 
taking the expectation value above requires us to evaluate
$\langle h_A^*(f,\hat{\Omega})h_{A'}(f',\hat{\Omega}')\rangle$.  The assumptions
that the stochastic background is stationary, unpolarized and isotropic lead us
to take
\beq
\langle h_A^*(f,\hat{\Omega})h_{A'}(f',\hat{\Omega}')\rangle = 
\delta^2(\hat{\Omega},\hat{\Omega}')\delta_{AA'}
\delta(f-f')H(f),
\label{hev}
\eeq
where $H(f)=H(-f)$ is the gravitational wave spectrum.  $H(f)$ is related to 
$\Omega_{\rm gw}(f)$ through
\beq
 \Omega_{\rm gw}(f)\equiv \frac{1}{\rho_{\rm crit}}\frac{d\rho_{\rm gw}}{d\ln f},
\label{omdef}
\eeq
where $\rho_{\rm crit}=8\pi/3H_0^2$ and
\beq
\rho_{\rm gw} = \frac{1}{32\pi}\langle \dot{h}_{ab}(t,\vec{x})
\dot{h}^{ab}(t,\vec{x})\rangle,
\label{eqrho}
\eeq
is the energy density in gravitational waves.  It follows from the plane
wave expansion \Deqn{pwexp} along with \Deqns{hev} and~(\ref{omdef}) in 
\Deqn{eqrho} that
\beq
H(f)=\frac{3H_0^2}{32\pi^3}|f|^{-3}\Omega_{\rm gw}(|f|),
\eeq
and therefore
\bea
\langle h_A^*(f,\hat{\Omega})h_{A'}(f',\hat{\Omega}')\rangle &=& 
\frac{3H_0^2}{32\pi^3}\delta^2(\hat{\Omega},\hat{\Omega}')\delta_{AA'}
\delta(f-f')\nonumber\\
&\pheq&\times|f|^{-3}\Omega_{\rm gw}(|f|),
\label{hevom}
\eea
which is sometimes written in terms of the characteristic strain
\beq
h_c^2(f)=\frac{3H_0^2}{2\pi^2}\frac{1}{f^2}\Omega_{\rm gw}(|f|).
\eeq
The expectation value we set out to evaluate is then
\beq\label{e:zexpect}
\langle\tilde{z}_1^*(f)\tilde{z}_2(f')\rangle
= \frac{3H_0^2}{32\pi^3}\frac{1}{\beta}\delta(f-f')|f|^{-3}
\Omega_{\rm gw}(|f|) \Gamma(|f|),
\eeq
where we defined
\bea
\Gamma(|f|) = &\beta& \sum_A \int_{S^2}d\hat{\Omega}\,
\left(e^{i2\pi f L_1(1+\hat{\Omega}\cdot\hat{p}_1)}-1\right)
\nonumber
\\
&\times&
\left(e^{-i2\pi f L_2(1+\hat{\Omega}\cdot\hat{p}_2)}-1\right)
F_1^A(\hat{\Omega})F_2^{A}(\hat{\Omega}),\nonumber\\
\label{orf}
\eea
the pulsar timing analogue of the overlap reduction function~\cite{AllenandRomano}, 
which has a normalization
factor $\beta$.  The normalization is chosen so that $\Gamma(|f|)=1$ for 
coincident, co-aligned detectors.  As we show below,
pulsar timing experiments are in a regime where the exponential
factors in \Deqn{orf} can be neglected.  In this situation, which we 
will assume henceforth, the
normalization factor is easy to determine and we 
have that
\bea
\Gamma_0 &\equiv& 
\frac{3}{4\pi} \sum_A\int_{S^2}d\hat{\Omega}\,
F_1^A(\hat{\Omega})F_2^{A}(\hat{\Omega})
\nonumber
\\
&=&3 \left\{ \frac{1}{3} + \frac{1-\cos\xi}{2}\left[ \ln\left(
    \frac{1-\cos\xi}{2}\right)  -\frac{1}{6}\right]\right\},
\nonumber
\\
\label{orfapprox}
\eea
where $\xi=\cos^{-1}(\hat p_1 \cdot \hat p_2)$ is the angle between the two 
pulsars.  This quantity is proportional to the Hellings and Downs 
curve~\cite{HD}. A detailed derivation of this result is provided for
completeness in Appendix~\ref{app:HDC}.

\begin{figure}[t]
\includegraphics[width=3.7in]{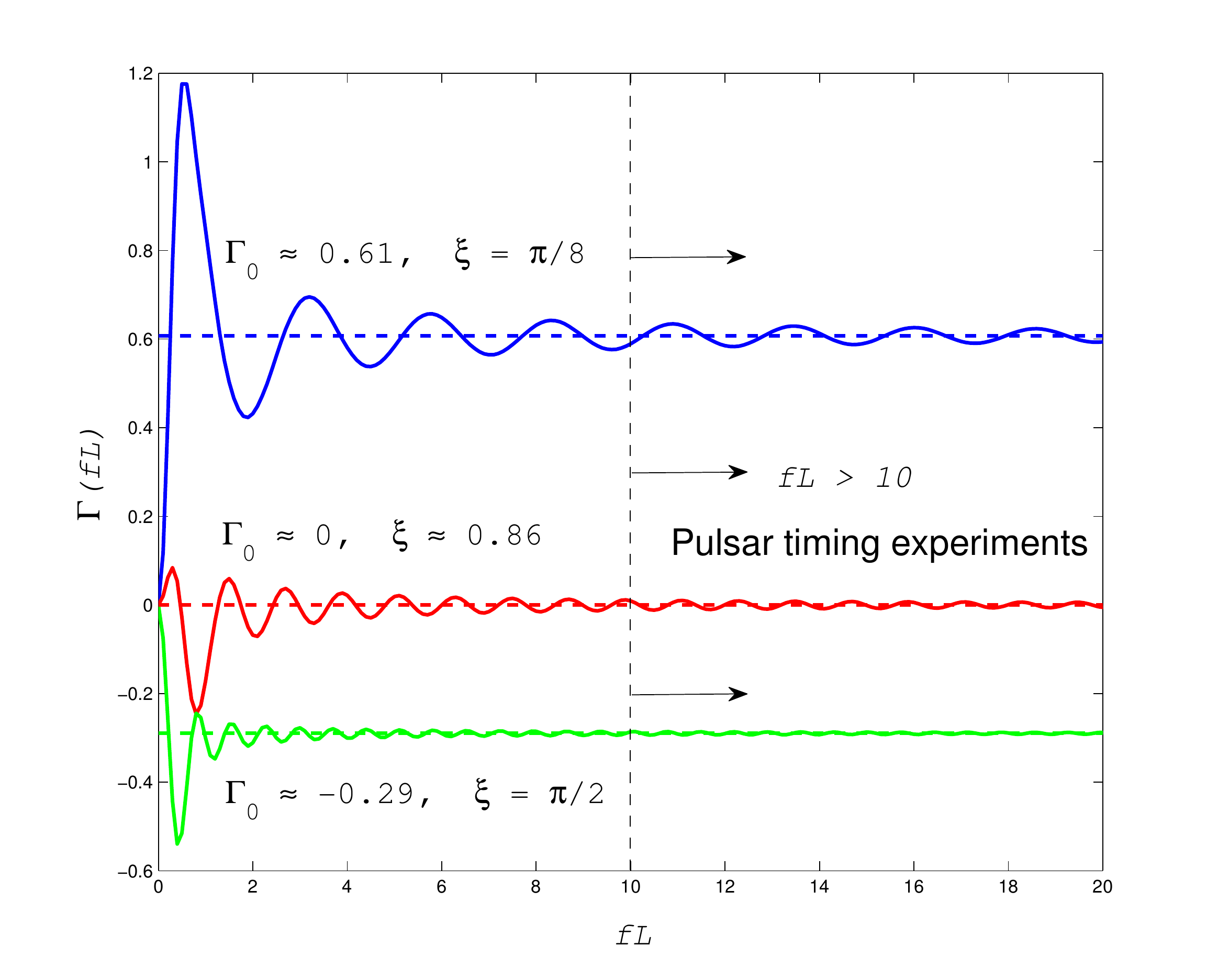}
\caption{Plot of the full overlap reduction, \Deqn{orf}, along
  with the approximation \Deqn{orfapprox} for two pulsars a distance $L$ from the
solar-system
barycenter. The overlap reduction function is a function of $fL$.  The top
two (blue) curves show \Deqn{orf} with $\beta = 3/4\pi$ (solid line)
and \Deqn{orfapprox} (dashed line) for two pulsars at an angle
$\xi=\pi/8$ as a function of $fL$. The middle (red) and bottom (green)
curves show the same quantities for two pulsars at $\xi \approx 0.86$
and $\xi=\pi/2$ respectively.  The smallest value of the 
frequency $f_{\rm min} \sim 0.1$~yr$^{-1}$ 
and the closest pulsars used in timing experiments are at a distance 
of $L_{\rm min} \sim 100$~ly so that $fL \gtrsim 10$. This 
range of $fL$ puts pulsar timing experiments in the regime
where \Deqn{orfapprox} is an excellent approximation to \Deqn{orf} and
we are justified in throwing out the pulsar term while remaining close to 
optimal.}
\label{fig:ORF}
\end{figure}

The rationale given in the literature for throwing out the pulsar term
in \Deqn{redshift}, or equivalently Eqs.~(\ref{eqzsom}) and
(\ref{delhdef}), is that the unknown metric perturbation at the
pulsars can be thought of as a kind of noise term which averages to
zero when performing a correlation between different pulsars. The
equivalent procedure in the frequency domain is to neglect the phase
factors in \Deqn{zredshift1}, or in terms of our optimal filter,
approximating \Deqn{orf} with \Deqn{orfapprox}.  The regime where the
approximate \Deqn{orfapprox} is valid, is helpful in quantifying the
accuracy of the rationale.  Figure~\ref{fig:ORF} shows the overlap
reduction function for two pulsars a distance $L$ from the solar-system
barycenter.
Since the distance to both pulsars is the same, the overlap reduction
function can be written as just a function of $fL$.  The top two
(blue) curves show \Deqn{orf} with $\beta = 3/4\pi$ (solid line) and
\Deqn{orfapprox} (dashed line) for two pulsars at an angle $\xi=\pi/8$
as a function of $fL$. The middle (red) and bottom (green) curves show
the same quantities for two pulsars at $\xi \approx 0.86$ and
$\xi=\pi/2$ respectively.  As discussed in the last section the
smallest value of the frequency $f_{\rm min} \sim 0.1$~yr$^{-1}$ and
the closest pulsars used in timing experiments are at a distance of
$L_{\rm min} \sim 100$~ly so that $fL \gtrsim 10$. As shown in
Fig.~\ref{fig:ORF} this range of $fL$ puts pulsar timing experiments
in the regime where \Deqn{orfapprox} is an excellent approximation to
\Deqn{orf}, and we can neglect the pulsar term while remaining close
to optimal.

Returning to \Deqn{eqmu}, we now have
\bea
\mu &=& \frac{3H_0^2}{32\pi^3}\frac{1}{\beta}T\int_{-\infty}^{\infty} df\, 
|f|^{-3}\Omega_{\rm gw}(|f|) \Gamma(|f|)\nonumber\\
&\approx& \frac{H_0^2}{8\pi^2}T\,\Gamma_0\int_{-\infty}^{\infty} df\, 
|f|^{-3}\Omega_{\rm gw}(|f|) .
\eea

With the assumption that the noise is much greater than the signal, the variance,
$\sigma^2$, depends only on the statistical properties of the noise in each 
detector. We have
\bea
\sigma^2
&\equiv&
\langle S^2\rangle-\langle S\rangle^2\approx \langle S^2\rangle
\nonumber
\\
&\approx& \frac{T}{4}\int_{-\infty}^{\infty} df\, P_1(|f|)P_2(|f|)
\left|\tilde{Q}(f)\right|^2
\label{eqsig}
\eea
where
\beq\label{e:psd}
\langle \tilde{n}_i^*(f)\tilde{n}_i(f')\rangle = \frac{1}{2}\delta(f-f') P_i(|f|),
\eeq
is the (one-sided) noise power spectrum.

With $\mu$ and $\sigma^2$ in hand, we next define a positive-definite inner 
product using the noise power spectra of the two detectors
\beq
(A,B)\equiv\int_{-\infty}^{\infty}df\, A^*(f)B(f)P_1(|f|)P_2(|f|).
\eeq
With this definition it is easy to see that
\bea
\mu &\approx& \frac{H_0^2}{8\pi^2}T\,\left(\tilde{Q}, 
\frac{\Omega_{\rm gw}(|f|)\Gamma_0}{|f|^3P_1(|f|)P_2(|f|)}
\right),\label{optfiltmu}\\
\sigma^2 &\approx& \frac{T}{4}\left(\tilde{Q},\tilde{Q}\right),
\label{sigip}
\eea
from which it follows from the definition of SNR in \Deqn{snrdef} and
Schwartz's inequality that the optimal filter is given by
\beq
\tilde{Q}(f)=\chi\frac{\Omega_{\rm gw}(|f|)\Gamma_0}{|f|^3P_1(|f|)P_2(|f|)},
\label{optfilt}
\eeq
for some normalization constant, $\chi$.  Our primary interest will be in
stochastic backgrounds with power law spectra, 
$\Omega_{\rm gw}(f)=\Omega_\alpha f^\alpha$ (for constant $\Omega_\alpha$).  In 
that case the normalization constant for the optimal filter, 
$\tilde{Q}_\alpha(f)$, is chosen so that
\beq
\mu=\Omega_\alpha T_0,\label{normcon}
\eeq
where $T_0$ is some arbitrary constant with dimensions of time.  From 
\Deqn{optfiltmu} it follows that
\beq
\chi = \Omega_{\alpha}\frac{T_0}{T}\frac{8\pi^2}{H_0^2}
\left[
\int_{-\infty}^{\infty}df\,\frac{\Omega_{\rm gw}^2(|f|)\Gamma_0^2}{f^6P_1(|f|)P_2(|f|)} 
\right]^{-1}.
\eeq
Finally, we can compute
\beq
{\rm SNR} \approx \frac{H_0^2}{4\pi^2}T^{1/2}\left[
\int_{-\infty}^{\infty}df\,\frac{\Omega_{\rm gw}^2(|f|)\Gamma_0^2}{f^6P_1(|f|)P_2(|f|)} 
\right]^{1/2}.
\label{optfiltsnr}
\eeq

The differences between these results and those for interferometers can all be
traced to the differing overlap reduction function $\Gamma(f)\approx\Gamma_0$.
The normalization of $\Gamma_0$ means that the maximal SNR (for co-incident, co-aligned
detectors) is only $5/6$ of that obtainable from interferometers, assuming the
noise power spectra are the same in each case. 

To construct the optimal filter, \Deqn{optfilt}, the noise power
spectra for the two pulsars $P_1(|f|)$ and $P_2(|f|)$ must be
determined. These can
either be modeled, or measured with the methods described in Section~\ref{issues}.
Once constructed the optimal filter can be applied in the frequency
domain. Section~\ref{issues} gives a prescription for taking Fourier
transforms of unevenly sampled data. The optimal filter can also be
inverse Fourier transformed and the correlation performed in the
time domain. It is unclear which of these two methods is more robust and
the authors of~\cite{Lommenea} will explore the time-domain approach in 
detail.

\subsection{\label{pta} The pulsar timing array}

The question we would like to address in this section is: Given
redshift measurements from $N$ different pulsars (which each have a
different noise profile), what is the best way to combine those
measurements to produce the \emph{most constraining} upper limit?  One
can consider the cross-correlations between any even number of
detectors, but it has been shown~\cite{AllenandRomano,newguys} 
that the optimal choice is the combination of pairwise
cross-correlations.  As it turns out, the solution to this problem
also solves the problem of non-stationarity in the noise power spectra
over periods longer than the typical observation time, $T$.

First let 
\beq
{}^{(ij)}S_1,{}^{(ij)}S_2,\ldots,{}^{(ij)}S_{n_{ij}},
\eeq
be $n_{ij}$ measurements of the cross-correlation between the $i$-th and $j$-th
pulsar.   We will assume that each measurement is taken 
with an optimal filter normalized so that while searching for a background of the
form $\Omega_{\rm gw}(f)=\Omega_\alpha f^\alpha$,
\beq
\left\langle {}^{(ij)}S_k\right\rangle = \Omega_\alpha T_0\equiv\mu,
\label{mnormcon}
\eeq
where $T_0$ is an arbitrary constant introduced for dimensional
reasons.  Each measurement therefore has the form
\bea
{}^{(ij)}S_k&=&\int_{-\infty}^{\infty}df\, \int_{-\infty}^{\infty}df'\, \delta_T(f-f')
\tilde{s}_{i,k}^*(f)\tilde{s}_{j,k}(f')\,{}^{(ij)}  Q_k(f)\nonumber\\
\label{optstatkij}
\eea
with
\bea
{}^{(ij)}  Q_k(f)={}^{(ij)}\chi_k\frac{\Omega_{\rm gw}(|f|)
{}^{(ij)}\Gamma_0}{|f|^3P_{i,k}(|f|)P_{j,k}(|f|)}.
\label{Qkij}
\eea
where ${}^{(ij)}\Gamma_0$ is the overlap reduction function of the $(ij)$
pulsar pair, $s_{i,k}$ is the $k$-th measurement of the signal from the $i$-th pulsar, and
$P_{i,k}$ is the associated noise power spectrum.  Additionally,
\beq
{}^{(ij)}\chi_k = \Omega_{\alpha}\frac{T_0}{{}^{(ij)}T_k}\frac{8\pi^2}{H_0^2}
\left[
\int_{-\infty}^{\infty}df\,\frac{\Omega_{\rm gw}^2(|f|){}^{(ij)}\Gamma_0^2}{f^6P_{i,k}(|f|)P_{j,k}(|f|)} 
\right]^{-1},
\label{normchi}
\eeq
where ${}^{(ij)}T_k$ is the observation time of the $k$-th measurement 
of the $(ij)$ pulsar pair.
Our task is to combine the ${}^{(ij)}S_k$ in a way that optimizes SNR.
The first step is to form the sample mean for each set of measurements
\beq
{}^{(ij)}\hat{\mu}\equiv\frac{1}{n_{ij}}\sum_{k=1}^{n_{ij}}{}^{(ij)}S_k,
\eeq
which is both an unbiased estimator and random variable. It therefore has a mean
\beq
\mu_{ij}\equiv\langle{}^{(ij)}\hat{\mu}\rangle=\mu,\label{muijsame}
\eeq
and a variance
\beq
\sigma_{ij}^2\equiv\langle{}^{(ij)}\hat{\mu}^2\rangle
-\langle{}^{(ij)}\hat{\mu}\rangle^2 = \frac{{}^{(ij)}\sigma^2}{n_{ij}},
\eeq
where
\beq
{}^{(ij)}\sigma^2= \sum_{k=1}^{n_{ij}}\frac{{}^{(ij)}T_k}{4}\int_{-\infty}^{\infty}
df\, {}^{(ij)}\chi_k^2 \frac{\Omega_{\rm gw}^2(|f|){}^{(ij)}\Gamma_0^2}{f^6P_{i,k}(|f|)P_{j,k}(|f|)}.
\label{varsopt}
\eeq
The next step is combine the sample mean for each set of measurements into a 
single estimator we can use to determine an upper bound on $\Omega_\alpha$ and
hence $\Omega_{\rm gw}=\Omega_\alpha f^\alpha$.  We do so by introducing an 
unbiased estimator consisting of a weighted average of the sample means
\beq
\hat{\mu}\equiv\frac{\displaystyle\sum_{i=1}^{l}\displaystyle\sum_{j<i}^{l}
\lambda_{ij}{}^{(ij)}\hat{\mu}}{
\displaystyle\sum_{i=1}^l\displaystyle\sum_{j<i}^l\lambda_{ij}},\label{wavg}
\eeq
for some constants, $\lambda_{ij}$, which has mean
\beq
\mu_{\hat{\mu}}\equiv\langle\hat{\mu}\rangle=\mu,
\eeq
and variance
\beq
\sigma^2_{\hat{\mu}}\equiv\langle\hat{\mu}^2\rangle-\langle\hat{\mu}\rangle^2=\frac{
\displaystyle\sum_{i=1}^{l}\displaystyle\sum_{j<i}^{l}\lambda_{ij}^2\sigma_{ij}^2}{
\left(\displaystyle\sum_{i=1}^{l}\displaystyle\sum_{j<i}^{l}\lambda_{ij}\right)^2}.
\label{varianceSopt}
\eeq
The object is to now determine the $\lambda_{ij}$ that maximizes the SNR of 
$\hat{\mu}$.  The (squared) SNR of $\hat{\mu}$ is
\beq
{\rm SNR}^2_{\hat{\mu}}\equiv\mu^2
\frac{\left(\displaystyle\sum_{i=1}^{l}\displaystyle\sum_{j<i}^{l}\lambda_{ij}\right)^2}{
\displaystyle\sum_{i=1}^{l}\displaystyle\sum_{j<i}^{l}\lambda_{ij}^2\sigma_{ij}^2}.
\eeq
To find the $\lambda_{ij}$ that maximize the SNR, we exploit the same trick 
that led us to the optimal filter.  Namely, we introduce an inner product
\beq
(A,B)\equiv \displaystyle\sum_{i=1}^{l}\displaystyle\sum_{j<i}^{l}A_{ij}^*B_{ij}\sigma_{ij}^2,
\label{mdip}
\eeq
which allows us to write
\beq
{\rm SNR}^2_{\hat{\mu}}\equiv\mu^2\frac{(\lambda,\sigma^{-2})}{(\lambda,\lambda)},
\eeq
from which it follows that choosing
\beq
\lambda_{ij}\propto\sigma_{ij}^{-2},
\eeq
maximizes the SNR.  The optimal statistic, 
choosing $\lambda_{ij}=\sigma_{ij}^{-2}$, is then given by
\bea\label{e:optimalstat}
S_{\rm opt} &=& 
\frac{\displaystyle\sum_{i=1}^{l}\displaystyle\sum_{j<i}^{l}
\sigma^{-2}_{ij}{}^{(ij)}\hat{\mu}}{
\displaystyle\sum_{i=1}^l\displaystyle\sum_{j<i}^l\sigma^{-2}_{ij}}\nonumber\\
&=&
\frac{\displaystyle\sum_{i=1}^{l}\displaystyle\sum_{j<i}^{l}{}^{(ij)}\sigma^{-2}
\displaystyle\sum_{k=1}^{n_{ij}}{}^{(ij)}S_k}{
\displaystyle\sum_{i=1}^l\displaystyle\sum_{j<i}^ln_{ij}{}^{(ij)}\sigma^{-2}}.
\label{Sopt}
\eea
Because the estimator defined in \Deqn{wavg} is unbiased and defined so that 
$\mu = \langle S_{\rm opt} \rangle=\Omega_\alpha T_0$, the estimate of 
$\hat \Omega_\alpha$ is found using
\beq
\hat \Omega_\alpha =\frac{\hat S_{\rm opt} }{T_0},
\label{omegapoint}
\eeq
where $\hat S_{\rm opt}$ is the measured value of the optimal statistic.  
The expected variance of $\hat S_{\rm opt}$ follows from \Deqn{varianceSopt},
\beq
\sigma^2_{\hat{\mu}} = 
\left( \displaystyle\sum_{i=1}^{l}\displaystyle\sum_{j<i}^{l}\sigma_{ij}^{-2}
\right)^{-1}.
\label{varianceSopt2}
\eeq

Aside from maximizing the SNR, the linear combination of sample means that forms
the optimal statistic in \Deqn{Sopt} serves two important and related purposes.
First of all, as mentioned at the beginning of this section, 
weighing each ${}^{(ij)}\hat{\mu}$ by the inverse of the squared 
variance means that less noisy measurements (those with smaller variances) 
contribute more  to the sum, which helps minimize the effect of long term 
non-stationarity.  This is augmented by the normalization convention
we chose in \Deqn{mnormcon} for the mean of each measurement.  Using 
\Deqn{sigip} and \Deqn{optfilt} with the $\lambda$ that follows from 
\Deqn{mnormcon}, we see that ${}^{(ij)}\sigma^{-2} \propto {}^{(ij)}T$,
and so longer observation times are also favored in the sum.

\subsection{\label{prescription}Computational procedure}

In this subsection we describe how the quantities necessary for a
stochastic background search are computed. The goal is to produce a
measurement of the optimal statistic, $\hat S_{\rm opt}$, using
\Deqn{Sopt}. The optimal statistic can then be used to make detection
or upper limit statements (see Section~\ref{ULDsec}).

First the power spectra spectra for each pulsar (and each stretch),
$P_{i,k}(|f|)$, must be determined. The spectra can
either be modeled or measured with the methods described in
Section~\ref{issues}.  Then the overlap reduction functions,
${}^{(ij)}\Gamma_0$, need to be computed for each pulsar pair. 
To optimize the statistic for particular spectra
the value of $\alpha$ (in $\Omega_{\rm gw}(f) =\Omega_\alpha
f^\alpha$) needs to be chosen. 
The normalizations,
${}^{(ij)}\chi_k$, can
then be computed using \Deqn{normchi}.  The normalizations allow us to
compute the variances, ${}^{(ij)}\sigma^{-2}$, given by \Deqn{varsopt},
in the numerator and denominator of \Deqn{Sopt}, as well as the
filters, ${}^{(ij)}Q_k(f)$, through \Deqn{Qkij}. 
Note that the unknown factors of $\Omega_\alpha$
cancel everywhere: From \Deqn{normchi} it is easy to see the
normalization ${}^{(ij)}\chi_k \propto \Omega_\alpha^{-1}$, so there
is a cancellation a factor of $\Omega_\alpha$ in \Deqn{Qkij}, and a
factor of $\Omega_\alpha^2$ in \Deqn{varsopt}.  
With these quantities in hand the cross-correlations, ${}^{(ij)}S_k$,
in \Deqn{optstatkij} can be computed by taking Fourier transforms of
the data (see Section~\ref{issues}). Alternatively, a set of
time-domain filters, ${}^{(ij)}Q_k(t)$, can be created by taking
inverse Fourier transforms of \Deqn{Qkij} and applied to the data in
the time domain using \Deqn{eqStd}.

Note that there is no dependence on the arbitrary constant $T_0$
introduced in \Deqn{mnormcon} for dimensional reasons. The
${}^{(ij)}\chi_k$ are linear in $T_0$ and enter
the variances quadratically (see \Deqn{varsopt}). The dependence
cancels in \Deqn{Sopt} because it is present in both numerator and
denominator.  $T_0$ also enters $S_{\rm
  opt}$ linearly through ${}^{(ij)}\chi_k$ in ${}^{(ij)}Q_k$
but cancels in \Deqn{omegapoint} so that the point estimate of $\hat
\Omega_\alpha$ is independent of $T_0$.

\subsection{\label{lhood}Likelihood approach}

The detection statistic that has been derived is also an optimal
statistic in the sense that it is the logarithm of the likelihood
ratio, at least in the limit where the expected signal is smaller than
the noise, and therefore it is the optimal statistic in both the
Bayesian sense and by the Neyman-Pearson criterion.  This section
is based on the likelihood analysis of~\cite{Allen:2002jw},
generalized to consider multiple detector pairs.

As we did previously, 
we assume that the noise is stationary and Gaussian, as
is the stochastic background.  For any given pulsar $i$ we assume that there
are discrete samples of data which forms a vector ${\mathbf{s}}_i$.
Although the discussion below does not place requirements on the data
sampling, we will assume that the observations of the pulsars all involve
the same number of points $N$ at the same evenly spaced sampling interval
so that sample $j$ of pulsar $i$ is ${\mathbf{s}}_i[j]=s_i(j\Delta t)$ where
$\Delta t$ is the sampling interval.  This signal vector is the sum
of a noise vector ${\mathbf{n}}_i$ and the redshift vector ${\mathbf{z}}_i$,
${\mathbf{s}}_i={\mathbf{z}}_i+{\mathbf{n}}_i$.
The data is a combination of two random processes: the instrumental noise
and the contribution from the stochastic background.  The autocorrelation
matrix
${\mathbf{R}}_i=\langle{\mathbf{s}}_i^\dagger\otimes{\mathbf{s}}_i\rangle$
is an $N\times N$ matrix which contains both of these contributions and, since
we assume Gaussian noise and stochastic background, this matrix completely
characterizes the distribution of the data.  As we did previously, 
we assume that the measurement
noise in a pulsar observation is independent of the noise in the observations
of other pulsars; the stochastic background, however, is correlated amongst
the pulsar signals.  This correlation is characterized by the stochastic
background correlation matrix
$\epsilon^2{\mathbf{S}}_{ij}=\langle{\mathbf{z}}_i^\dagger\otimes{\mathbf{z}}_j\rangle$.
Here $\epsilon$ is an order parameter which we will use to expand the
probability distribution in powers of the small stochastic background signal.
It can also be interpreted as an overall amplitude parameter of the stochastic
background.
The probability distribution for the collection of all pulsar observations
is given by a multidimensional Gaussian distribution
\begin{equation}
  p({\mathbf{x}}|\epsilon) =
  \frac{1}{\sqrt{\det(2\pi{\boldsymbol\Sigma})}}
  \exp\left(-{\textstyle\frac12}{\mathbf{x}}^\dagger\cdot
  {\boldsymbol\Sigma}^{-1}\cdot{\mathbf{x}}\right)
\end{equation}
where
\begin{equation}
  {\mathbf{x}}=\left[
  \begin{array}{c}
    {\mathbf{s}}_1 \\
    {\mathbf{s}}_2 \\
    \vdots \\
    {\mathbf{s}}_l
  \end{array}
  \right]
\end{equation}
is a column vector formed from all of the data vectors and
\begin{equation}
  {\boldsymbol\Sigma} = \left[
  \begin{array}{cccc}
    {\mathbf{R}}_1 & \epsilon^2 {\mathbf{S}}_{12} & \cdots
      & \epsilon^2 {\mathbf{S}}_{1l} \\
    \epsilon^2 {\mathbf{S}}_{21} & {\mathbf{R}}_{2} & \cdots
      & \epsilon^2 {\mathbf{S}}_{2l} \\
    \vdots & \vdots & \ddots & \vdots \\
    \epsilon^2 {\mathbf{S}}_{l1} & \epsilon^2 {\mathbf{S}}_{l2} & \cdots
      & {\mathbf{R}}_l 
  \end{array}
  \right]
\end{equation}
is the correlation matrix for the collective observation vector ${\mathbf{x}}$.
In this weak signal limit we find
\begin{widetext}
\begin{eqnarray}
  &&{\boldsymbol\Sigma}^{-1} = \left[
  \begin{array}{cccc}
    {\mathbf{R}}_1^{-1} & {\mathbf{0}} & \cdots & {\mathbf{0}}  \\
    {\mathbf{0}} & {\mathbf{R}}_{2}^{-1} & \cdots & {\mathbf{0}} \\
    \vdots & \vdots & \ddots & \vdots \\
    {\mathbf{0}} & {\mathbf{0}} & \cdots & {\mathbf{R}}_l^{-1} 
  \end{array}
  \right]
  - \epsilon^2 \left[
  \begin{array}{cccc}
    {\mathbf{0}}
      & {\mathbf{R}}_1^{-1}\cdot{\mathbf{S}}_{12}\cdot{\mathbf{R}}_2^{-1}
      & \cdots
      & {\mathbf{R}}_1^{-1}\cdot{\mathbf{S}}_{1l}\cdot{\mathbf{R}}_l^{-1} \\
    {\mathbf{R}}_2^{-1}\cdot{\mathbf{S}}_{21}\cdot{\mathbf{R}}_1^{-1}
      & {\mathbf{0}}
      & \cdots
      & {\mathbf{R}}_2^{-1}\cdot{\mathbf{S}}_{2l}\cdot{\mathbf{R}}_l^{-1} \\
    \vdots & \vdots & \ddots & \vdots \\
    {\mathbf{R}}_l^{-1}\cdot{\mathbf{S}}_{l1}\cdot{\mathbf{R}}_1^{-1}
      & {\mathbf{R}}_l^{-1}\cdot{\mathbf{S}}_{l2}\cdot{\mathbf{R}}_2^{-1}
      & \cdots
      & {\mathbf{0}}
  \end{array}
  \right] \nonumber\\
  && + \epsilon^4\left[
  \begin{array}{cccc}
    {\displaystyle\sum_{\stackrel{\scriptstyle m=1}{m\ne1}}^l}{\mathbf{R}}_1^{-1}\cdot{\mathbf{S}}_{1m}\cdot{\mathbf{R}}_m^{-1}\cdot{\mathbf{S}}_{m1}\cdot{\mathbf{R}}_1^{-1}
      & {\displaystyle\sum_{\stackrel{\scriptstyle m=1}{m\ne1,2}}^l}{\mathbf{R}}_1^{-1}\cdot{\mathbf{S}}_{1m}\cdot{\mathbf{R}}_m^{-1}\cdot{\mathbf{S}}_{m2}\cdot{\mathbf{R}}_2^{-1}
      & \cdots 
      & {\displaystyle\sum_{\stackrel{\scriptstyle m=1}{m\ne1,l}}^l}{\mathbf{R}}_1^{-1}\cdot{\mathbf{S}}_{1m}\cdot{\mathbf{R}}_m^{-1}\cdot{\mathbf{S}}_{ml}\cdot{\mathbf{R}}_l^{-1} \\
    {\displaystyle\sum_{\stackrel{\scriptstyle m=1}{m\ne2,1}}^l}{\mathbf{R}}_2^{-1}\cdot{\mathbf{S}}_{2m}\cdot{\mathbf{R}}_m^{-1}\cdot{\mathbf{S}}_{m1}\cdot{\mathbf{R}}_1^{-1}
      & {\displaystyle\sum_{\stackrel{\scriptstyle m=1}{m\ne2}}^l}{\mathbf{R}}_2^{-1}\cdot{\mathbf{S}}_{2m}\cdot{\mathbf{R}}_m^{-1}\cdot{\mathbf{S}}_{m2}\cdot{\mathbf{R}}_2^{-1}
      & \cdots 
      & {\displaystyle\sum_{\stackrel{\scriptstyle m=1}{m\ne2,l}}^l}{\mathbf{R}}_2^{-1}\cdot{\mathbf{S}}_{2m}\cdot{\mathbf{R}}_m^{-1}\cdot{\mathbf{S}}_{ml}\cdot{\mathbf{R}}_l^{-1} \\
    \vdots & \vdots & \ddots & \vdots \\
    {\displaystyle\sum_{\stackrel{\scriptstyle m=l}{m\ne l,1}}^l}{\mathbf{R}}_l^{-1}\cdot{\mathbf{S}}_{lm}\cdot{\mathbf{R}}_m^{-1}\cdot{\mathbf{S}}_{m1}\cdot{\mathbf{R}}_1^{-1}
      & {\displaystyle\sum_{\stackrel{\scriptstyle m=l}{m\ne l,2}}^l}{\mathbf{R}}_l^{-1}\cdot{\mathbf{S}}_{lm}\cdot{\mathbf{R}}_m^{-1}\cdot{\mathbf{S}}_{m2}\cdot{\mathbf{R}}_2^{-1}
      & \cdots 
      & {\displaystyle\sum_{\stackrel{\scriptstyle m=l}{m\ne l}}^l}{\mathbf{R}}_l^{-1}\cdot{\mathbf{S}}_{lm}\cdot{\mathbf{R}}_m^{-1}\cdot{\mathbf{S}}_{ml}\cdot{\mathbf{R}}_l^{-1}
  \end{array}
  \right] \nonumber\\
  &&+ O(\epsilon^6)
\end{eqnarray}
and
\begin{equation}
  \ln\det{\boldsymbol\Sigma} = \sum_{i=1}^l \ln\det{\mathbf{R}}_i
  + \epsilon^4 \sum_{i=1}^l\sum_{j<i}^l \tr({\mathbf{R}}_i^{-1}\cdot{\mathbf{S}}_{ij}\cdot{\mathbf{R}}_j^{-1}\cdot{\mathbf{S}}_{ji})
  + O(\epsilon^6).
\end{equation}
The logarithm of the likelihood ratio is
\begin{eqnarray}\label{e:likelihood}
  \ln\Lambda&=&\ln p({\mathbf{x}}|\epsilon) - \ln p({\mathbf{x}}|0) \nonumber\\
  &=& \epsilon^2 \sum_{i=1}^l \sum_{j<i}^l \Re\left(
  {\mathbf{s}}_i^\dagger\cdot
  {\mathbf{R}}_i^{-1}\cdot{\mathbf{S}}_{ij}\cdot
  {\mathbf{R}}_j^{-1}\cdot{\mathbf{s}}_j
  \right) \nonumber\\
  && + \frac{1}{2}\epsilon^4 \sum_{i=1}^l \left\{ \sum_{j<i}^l \tr({\mathbf{R}}_i^{-1}\cdot{\mathbf{S}}_{ij}\cdot{\mathbf{R}}_j^{-1}\cdot{\mathbf{S}}_{ji})
  - 2 \sum_{j\le i}^l \sum_{\stackrel{\scriptstyle m=1}{m\ne i,j}}^l \Re\left( {\mathbf{s}}_i^\dagger\cdot{\mathbf{R}}_i^{-1}\cdot{\mathbf{S}}_{im}\cdot{\mathbf{R}}_m^{-1}\cdot{\mathbf{S}}_{mj}\cdot{\mathbf{R}}_j^{-1}\cdot{\mathbf{s}}_j \right)\right\}  \nonumber\\
 && + O(\epsilon^6) \nonumber\\
 &=& \epsilon^2 {\mathcal{S}} - {\textstyle\frac12}\epsilon^4 {\mathcal{N}}^2 + O(\epsilon^6).
\end{eqnarray}
\end{widetext}
This is the optimal detection statistic for a weak stochastic background.
We have identified ${\mathcal{S}}$ as the $O(\epsilon^2)$ term and
$-2{\mathcal{N}}^2$ as the $O(\epsilon^4)$ term of the log-likelihood ratio.

The \emph{locally optimal} detection statistic is obtained in the
$\epsilon\to0$ limit; it is the leading $O(\epsilon^2)$ term:
\begin{equation}\label{e:locallyoptimal}
  \lim_{\epsilon\to0} \frac{\ln\Lambda}{\epsilon^2} = {\mathcal{S}} =
  \sum_{i=1}^l \sum_{j<i}^l \Re\left(
  {\mathbf{s}}_i^\dagger\cdot
  {\mathbf{R}}_i^{-1}\cdot{\mathbf{S}}_{ij}\cdot
  {\mathbf{R}}_j^{-1}\cdot{\mathbf{s}}_j
  \right).
\end{equation}
Although this presentation has been described in terms of observational
vectors in the time-domain, the derivation of the likelihood ratio has not
explicitly required this choice of basis.  It is convenient to perform
a unitary transformation that diagonalizes the various correlation matrices.
This transformation is called a Karhunen-Loeve transformation; for a
stationary process with a correlation time much shorter than the time spanned
by the $l$ samples, the linear combinations of the time series that diagonalize
the correlation matrices asymptotically approach the discrete Fourier
transform.  Therefore we can approximately express our result in the frequency
domain where the ${\mathbf{R}}_i$ and ${\mathbf{S}}_{ij}$ matrices can be
understood in terms of the power spectrum and the expectation value of the
redshift cross-correlation respectively [cf. Eq.~(\ref{e:psd}) and
Eq.~(\ref{e:zexpect})].  The locally-optimal detection statistic is therefore
\begin{eqnarray}\label{e:locallyoptimalfd}
  {\mathcal{S}} &=& \frac{3H_0^2}{16\pi^3}\frac{1}{\beta}
  \sum_{i=1}^l \sum_{j<i}^l
  \int_{-\infty}^{\infty}
  \frac{\hat\Omega_{\mathrm{gw}}(|f|) \, {}^{(ij)}\Gamma(|f|)\tilde{s}_i^\ast(f)\tilde{s}_j(f)}
{f^3 P_i(|f|)P_j(|f|)}\,df \nonumber\\
  &=& \frac{1}{2}\hat\Omega_\alpha T_0 \sum_{i=1}^l \sum_{j<i}^l {}^{(ij)}\sigma^{-2}\,{}^{(ij)}S
\end{eqnarray}
where $\Omega_{\mathrm{gw}}(f)=\epsilon^2\hat\Omega_{\mathrm{gw}}(f)$ and
$\Omega_\alpha=\epsilon^2\hat\Omega_\alpha$.
This is the same optimal detection statistic $S_{\mathrm{opt}}$
of Eq.~(\ref{e:optimalstat}) (with the simplification of $n_{ij}=1$) up to a
normalization constant.

The locally optimal statistic is optimal in the limit of weak signals.
However, the likelihood ratio is only determined by this statistic up
to a unknown factor which depends on the (unknown) strength of the signal.
It is important now to distinguish between the assumed amplitude of the
stochastic background, $\epsilon$, and the true amplitude,
$\epsilon_{\mathrm{true}}$.  The true gravitational wave spectrum
$\Omega_{\mathrm{gw}}(f)$ is now related to the template spectrum
$\hat\Omega_{\mathrm{gw}}(f)$ via
$\Omega_{\mathrm{gw}}(f)=\epsilon_{\mathrm{true}}^2\hat\Omega_{\mathrm{gw}}$.
To measure the strength of the stochastic background given a set of pulsar
observations, we can use the maximum likelihood estimator: the value
of $\epsilon$, $\epsilon_{\mathrm{MLE}}$, for which the likelihood ratio
is a maximum.  That is, we wish to find the value of $\epsilon_{\mathrm{MLE}}$ 
for which $d\ln\Lambda/d\epsilon^2|_{\epsilon_{\mathrm{MLE}}}=0$.  From
Eq.~(\ref{e:likelihood}) we see that this estimate is
\begin{equation}\label{e:mle}
  \epsilon_{\mathrm{MLE}}^2 = {\mathcal{N}}^{-2}{\mathcal{S}}
\end{equation}
where ${\mathcal{N}}^2$, from the $O(\epsilon^4)$ term of the log-likelihood
ratio, is a normalizing factor which also includes the data.  By substituting
Eq.~(\ref{e:mle}) into Eq.~(\ref{e:likelihood}) we obtain the maximum
likelihood detection statistic
\begin{equation}
  \max_{\epsilon}\ln\Lambda \simeq \frac{1}{2}\frac{{\mathcal{S}}^2}{{\mathcal{N}}^2}
\end{equation}
where the terms of $O(\epsilon^6)$ have been discarded.
Notice that this statistic is not simply the square of the cross
correlation statistic.  The data also appears in the factor
${\mathcal{N}}^{-2}$.  This factor effectively suppresses elements of the
pulsar network where the data measured greatly exceeds the normal noise level.

Some insight into the maximum likelihood detection statistic and
the maximum likelihood amplitude estimate can be obtained by computing
the expectation value of the log-likelihood ratio, Eq.~(\ref{e:likelihood}).
We find, to leading order in $\epsilon$, 
\begin{eqnarray}
  \langle{\mathcal{N}}^2\rangle
  &=&
  \epsilon_{\mathrm{true}}^{-2}\langle{\mathcal{S}}\rangle
  \nonumber\\
  &=&
  \left(\frac{3H_0}{32\pi^3}\right)^2 \frac{1}{\beta^2} T
  \sum_{i=1}^l \sum_{j<i}^l
  \int_{-\infty}^{\infty} df\,\frac{\hat\Omega_{\mathrm{gw}}^2(|f|)\,{}^{(ij)}\Gamma^2(|f|)}{f^6P_i(|f|)P_j(|f|)}.
  \nonumber\\
\end{eqnarray}
Therefore
\begin{eqnarray}\label{e:expectlikelihood}
  \langle\ln\Lambda\rangle &=&
  \epsilon^2\langle{\mathcal{S}}\rangle
  - {\textstyle\frac12}\epsilon^4\langle{\mathcal{N}^2}\rangle \nonumber\\
  &=& \epsilon^2 (\epsilon^2_{\mathrm{true}} - {\textstyle\frac12}\epsilon^2)
  \left(\frac{3H_0}{32\pi^3}\right)^2 \frac{1}{\beta^2} T
  \nonumber\\
  && \times\sum_{i=1}^l \sum_{j<i}^l
  \int_{-\infty}^{\infty} df\,\frac{\hat\Omega_{\mathrm{gw}}^2(|f|)\,{}^{(ij)}\Gamma^2(|f|)}{f^6P_i(|f|)P_j(|f|)}
  \nonumber\\
  &&+ O(\epsilon^6).
\end{eqnarray}
If we ignore the $O(\epsilon^6)$ terms, this is maximized when
$\epsilon_{\mathrm{MLE}}=\epsilon_{\mathrm{true}}$, in which case
\begin{eqnarray}
  \max_\epsilon\langle\ln\Lambda\rangle &\simeq&
  \frac{1}{2} \left(\frac{3H_0}{32\pi^3}\right)^2 \frac{1}{\beta^2} T
  \nonumber\\
  && \times\sum_{i=1}^l \sum_{j<i}^l
  \int_{-\infty}^{\infty} df\,\frac{\Omega_{\mathrm{gw}}^2(|f|)\,{}^{(ij)}\Gamma^2(|f|)}{f^6P_i(|f|)P_j(|f|)}.
  \nonumber\\
\end{eqnarray}
This gives a scale of the value of the likelihood ratio we would expect to
achieve.

\section{\label{ULDsec}Upper limits and detection}

Several methods exist in the LIGO literature that are appropriate for
upper limit computation and detection using pulsar timing 
data~\cite{AllenandRomano,Abbott:2003hr,%
Abbott:2005ez,Abbott:2006zx,Abbott:2007tw,Abbott:2007wd}. These
methods can be divided into two classes: Frequentist and Bayesian.

We expect that the optimal statistic \Deqn{Sopt} will be formed 
from a large number of pulsar pairs. For example, the Parkes Pulsar Timing 
Array~\cite{Manchester:2007mx,PPTA} consists of 20 pulsars and the optimal 
statistic could be 
constructed from up to 190 cross-correlation pairs. In this case we can make use of
of the central limit theorem: The distribution of $\hat S_{\rm opt}$ should be 
well approximated  by a Gaussian with a mean 
$\mu = \langle S_{\rm opt} \rangle=\Omega_\alpha T_0$ and
variance $\sigma^2_{\hat{\mu}}$ given by \Deqn{varianceSopt2}, namely,
\bea
p( \hat S_{\rm opt} | \mu  \sigma_{\hat{\mu}})
 =\frac{1}{\sigma_{\hat{\mu}} \sqrt{2\pi} } \exp 
\left( \frac{-( \hat S_{\rm opt} - \mu )^2}{2 \sigma^2_{\hat{\mu}}} \right).
\eea

A straightforward frequentist upper limit can then be set by finding the 
value of $\mu_{\rm ul}$ such that in some pre-determined fraction $C$ 
(called the confidence) of hypothetical experiments, the value of the
optimal statistic exceeds the actual value $\hat S_{\rm opt}$ 
found in the search. In other words we would like to find the value
$\mu_{\rm ul}$ such that
\bea
\int_{\hat S_{\rm opt}}^{\infty} dS_{\rm opt} \,\, 
p( S_{\rm opt} | \mu_{\rm ul}  \sigma_{\hat{\mu}})=C.
\eea
The solution to this is
\bea
\mu_{\rm ul} = \hat S_{\rm opt} + \sqrt{2} \sigma_{\hat{\mu}} {\rm erfc}^{-1}(2(1-C)).
\label{eq:frequl}
\eea
The assertion is that the real value of $\mu$ is less than 
$\mu_{\rm ul}$ with confidence $C$, because if $\mu=\mu_{\rm ul}$, 
a fraction $C$ of the time we would have observed 
a value of $S_{\rm opt}$ greater than $\hat S_{\rm opt}$.
An equivalent, though potentially more robust, frequentist method to set upper limits
involves performing simulated signal injections in the timing data set. 
Multiple injections are performed to determine the value of $\mu_{\rm ul}$ 
such that a fraction $C$ of the time the value of the optimal statistic 
measured in the data sets with injections exceeds the value found in the search.
Frequentist detection methods such as 
Neyman-Pearson or maximum-likelihood are well described in the literature
(see, for example, \cite{AllenandRomano} and references 
therein) and we will not discuss them here. Additionally Feldman and 
Cousins~\cite{FeldCous} provide a means to smoothly transition 
between upper limits and detection.

Bayesian upper limits can be computed by constructing a posterior
distribution using the value of the optimal statistic found in the search,
and variance along with priors. We begin by applying the product rule 
to the probability density of $\mu$ along with the 
measured value $\hat S_{\rm opt}$ given $\sigma_{\hat{\mu}}$ to write,
\bea
p(\mu \hat S_{\rm opt}| \sigma_{\hat{\mu}}) &=& 
p(\mu| \hat S_{\rm opt} \sigma_{\hat{\mu}})p(\hat S_{\rm opt} |
\sigma_{\hat{\mu}})  
\nonumber
\\
&=&  p(\hat S_{\rm opt}| \mu \sigma_{\hat{\mu}})p(\mu |\sigma_{\hat{\mu}}),
\eea
then solve for $p(\mu| \hat S_{\rm opt} \sigma_{\hat{\mu}})$ to obtain Bayes' theorem, 
\bea 
p(\mu| \hat S_{\rm opt} \sigma_{\hat{\mu}})  =  p(\hat S_{\rm opt}| \mu
\sigma_{\hat{\mu}}) 
\frac{p(\mu |\sigma_{\hat{\mu}})}{p(\hat S_{\rm opt} |\sigma_{\hat{\mu}})},
\eea 
the posterior probability density for $\mu$, or equivalently $\Omega_\alpha$.
One can then choose a prior $p(\mu |\sigma_{\hat{\mu}})$ (for example requiring 
$\mu > 0$) and normalize the probability distribution (the probability 
$p(\hat S_{\rm opt}|\sigma_{\hat{\mu}})$ does not depend on $\mu$ so it is 
a prior dependent normalization constant), and find the
$\mu_{\rm ul}$ such that
\bea 
{\cal M} \int^{\mu_{\rm ul}} _{-\infty} d\mu \,
p(\mu| \hat S_{\rm opt} \sigma_{\hat{\mu}})  =  C,
\label{bayesianul}
\eea
where $\cal M$ is the normalization constant. For sufficiently simple choices of
the prior distribution $p(\mu |\sigma_{\hat{\mu}})$ 
the integral \Deqn{bayesianul} can be performed analytically to 
obtain the Bayesian analog of \Deqn{eq:frequl}. As with frequentist 
methods~\cite{AllenandRomano,FeldCous},  Bayesian detection methods involve 
selecting thresholds, in this case on the odds ratio, which is 
the ratio of the posteriors, suitably integrated over, say, different 
ranges of $\mu$. For more details see Refs.~\cite{loredo,gregoryandloredo}.

\begin{figure}[t]
\centering
\subfigure{\includegraphics[width=3.7in]{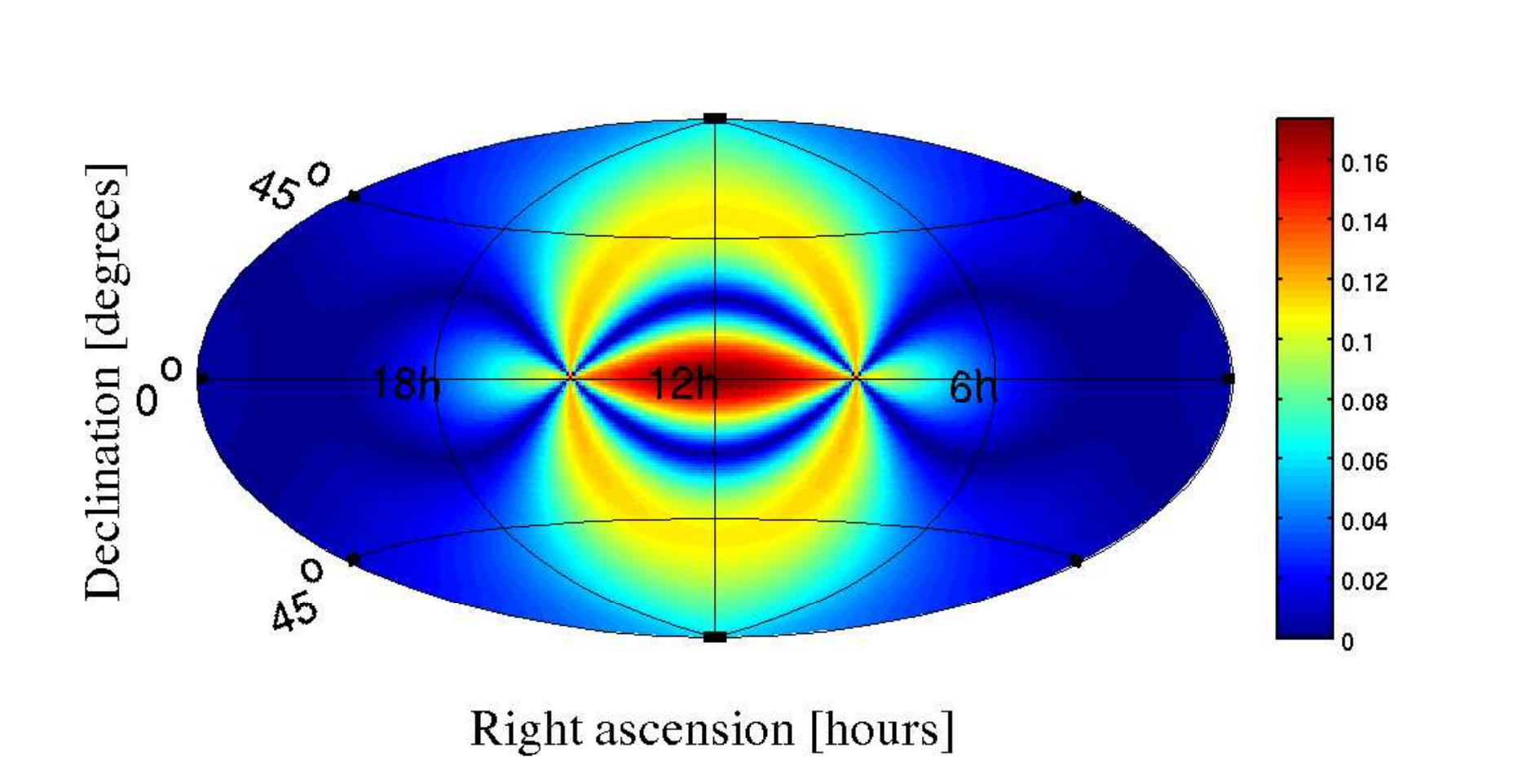}}\\
\subfigure{\includegraphics[width=3.7in]{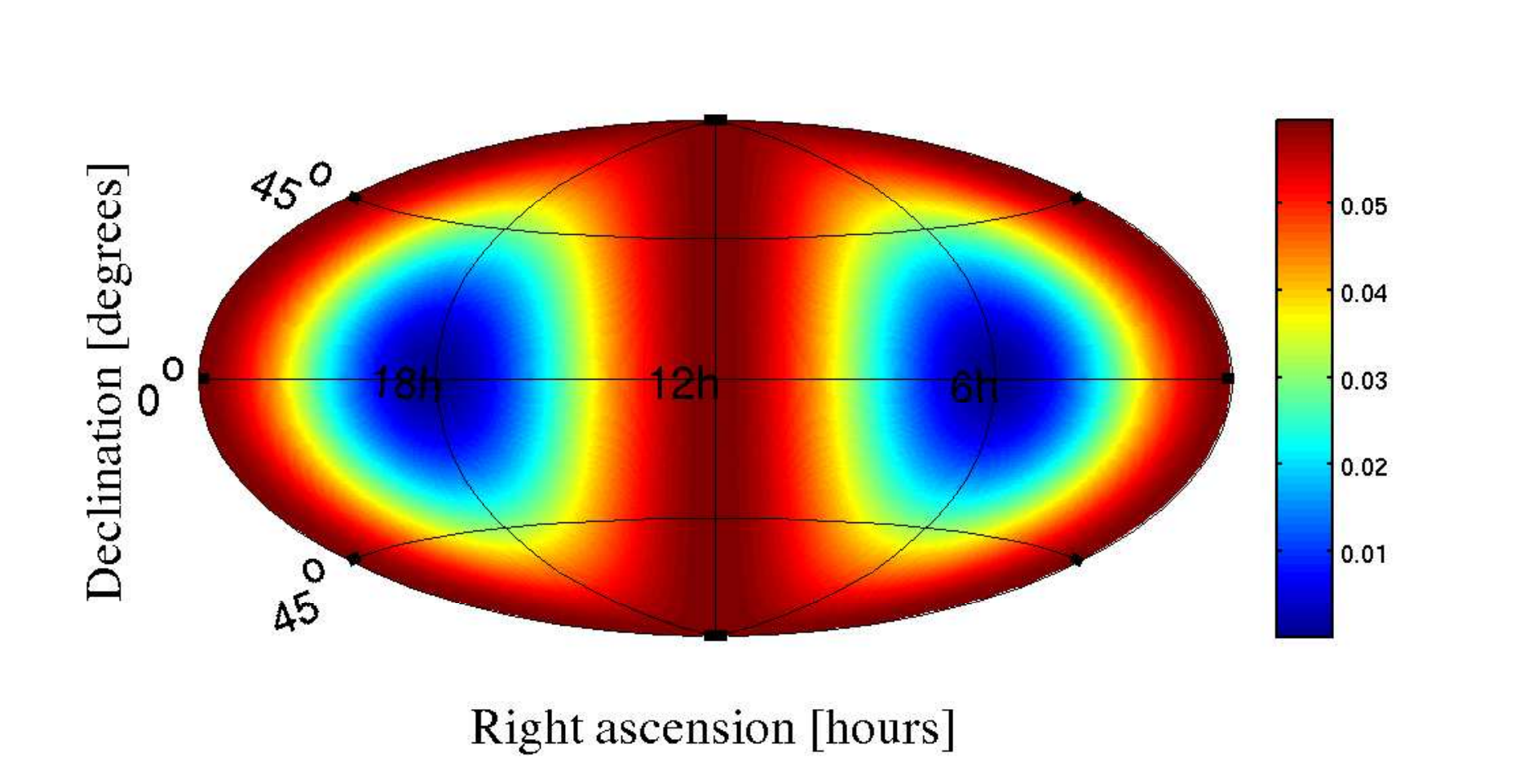}}
\caption{Plots of $|{}^{(ij)}\Gamma_{\hat \Omega}|$ from Eq.~(\ref{ORFdir}) for two pulsars 
with $\xi=\pi/2$ (top panel) and $\xi=\pi$ degrees (bottom panel)}
\label{fig:GWBeam}
\end{figure}

\section{\label{SMsec}A pulsar timing radiometer: Constructing skymaps of the stochastic background}

A skymap may be created by computing $\Omega_\text{gw}(f)$
for a collection of pixels in the sky. We do this by assuming
that the only signal present comes from a single location 
on the sky.  We begin by relaxing the assumption that the stochastic 
background is isotropic. That is, we take
\bea
\langle h_A^*(f,\hat{\Omega})h_{A'}(f',\hat{\Omega}')\rangle &=& 
\frac{3H_0^2}{32\pi^3}\delta^2(\hat{\Omega},\hat{\Omega}')\delta_{AA'}
\delta(f-f')\nonumber\\
&\pheq&\times P(\hat \Omega) |f|^{-3}\Omega_{\rm gw}(|f|),
\eea
where $P(\hat \Omega)$ is the strength or brightness~\cite{mitra} of
gravitational waves from the direction $\hat \Omega$.  

In this case, the overlap reduction function takes the modified form,
\bea
{}^{(ij)}\Gamma_{P} &=& 
\frac{3}{4\pi} \sum_A\int_{S^2}d\hat{\Omega}\, P(\hat \Omega)
F_i^A(\hat{\Omega})F_j^{A}(\hat{\Omega}).
\eea
where we've ignored the pulsar phase factors, and the optimal filter
is given by
\beq
{}^{(ij)}\tilde{Q}_P(f)={}^{(ij)}\chi\frac{{}^{(ij)}
\Gamma_{P}\Omega_{\rm gw}(|f|)}{|f|^3P_i(|f|)P_j(|f|)},
\label{optfiltS}
\eeq
where we have suppressed the $k$ index which specifies 
the particular measurement of the $(ij)$ pulsar pair.
We can further optimize for point sources by taking $P(\hat
\Omega)=\delta^2(\hat \Omega-\hat \Omega')$. The optimal filter then becomes,
\beq
{}^{(ij)}\tilde{Q}_{\hat \Omega}(f)={}^{(ij)}\chi\frac{{}^{(ij)}
\Gamma_{\hat \Omega}\Omega_{\rm gw}(|f|)}{|f|^3P_i(|f|)P_j(|f|)},
\label{optfiltS2}
\eeq
with, 
\bea
{}^{(ij)}\Gamma_{\hat \Omega} &=& 
\frac{3}{4\pi} \sum_A F_i^A(\hat{\Omega})F_j^{A}(\hat{\Omega}).
\label{ORFdir}
\eea
Figure~\ref{fig:GWBeam} shows two examples of the sky location
dependent overlap reduction function.  The top panel shows
$|\Gamma_{\hat \Omega}|$ from Eq.~(\ref{ORFdir}) for two pulsars with
$\xi=\pi/2$ located at $0^\circ$ Dec and 9h and 15h RA respectively.
The bottom panel shows the same quantity for two pulsars with
$\xi=\pi$ located at  $0^\circ$ Dec and 6h and 18h RA respectively.

One could also imagine computing the overlap reduction function for
each term in a multipole expansion of $P(\hat \Omega)$. The overlap
reduction function for the monopole term in the expansion (appropriate
for an isotropic stochastic gravitational wave search) is the
Hellings-Downs curve given by Eq.~(\ref{orfapprox}) in
Section~\ref{DSsec}. Surprisingly, the dipole overlap
reduction function is given by a similarly simple equation. We find
\begin{eqnarray}
  \Gamma_{\rm dip} &=& \left(\cos\alpha_1 + \cos\alpha_2\right)
\nonumber
\\
&\times& \left(2- \frac{3}{2}\cos\xi + 6\tan^2\frac{\xi}{2}
  \ln\left(\sin\frac{\xi}{2}\right)\right),
\end{eqnarray}
where as before $\xi$ is the angle between the two pulsars, and $\alpha_1$ and
$\alpha_2$ are the angles each of two pulsars make to the direction of
the dipole. A detailed derivation of this result is given in
Appendix~\ref{app:dipole}. This result is relevant to searches for a
dipole anisotropy in the gravitational wave sky using pulsar timing
data.

\begin{figure}[t]
\includegraphics[width=3.7in]{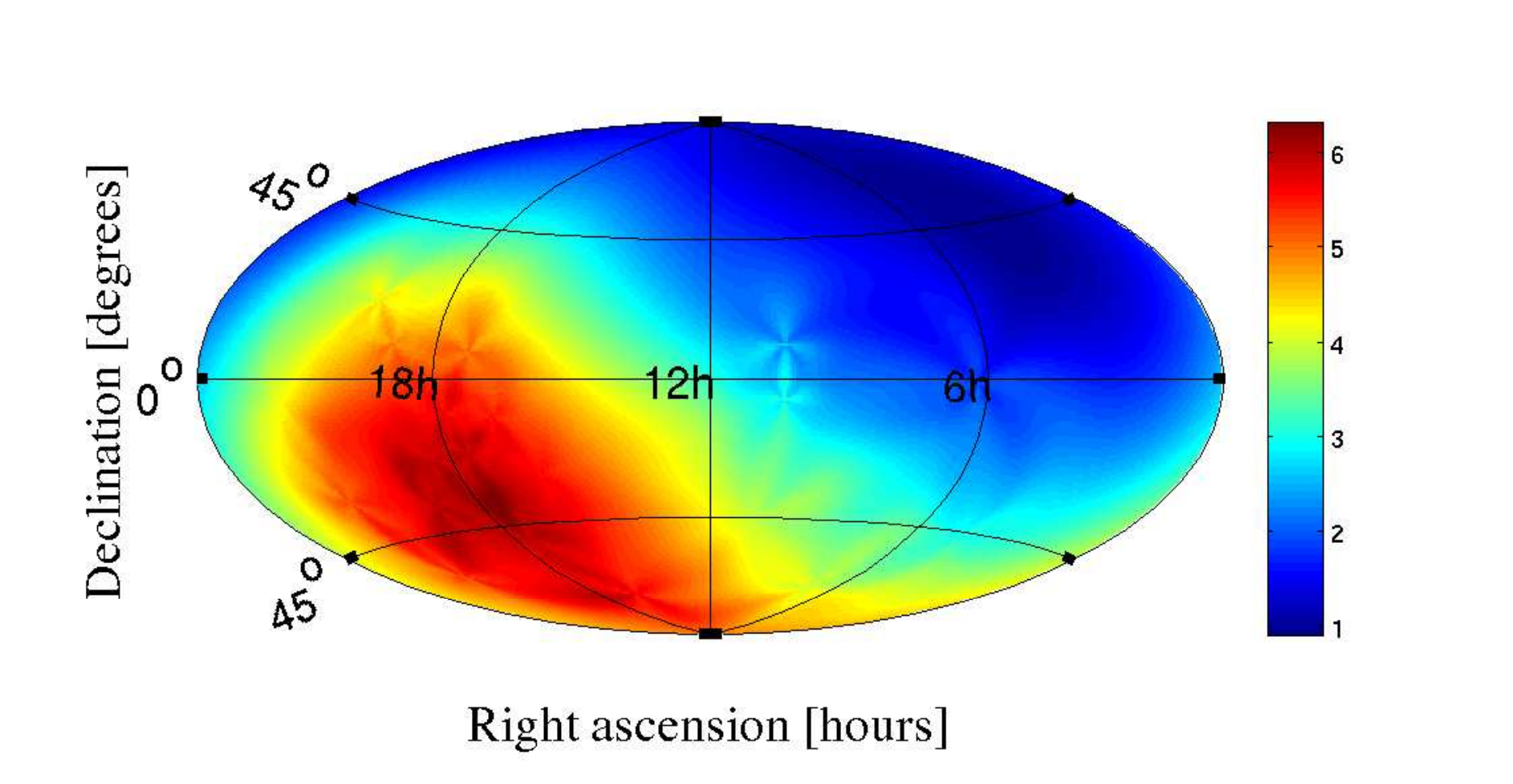}
\caption{Skymap of the sensitivity, \Deqn{avsnr}, for the Parkes Pulsar Timing
Array.}
\label{fig:parkessens}
\end{figure}

The sky-dependence of the sensitivity of a pulsar network can be estimated 
by computing the signal to noise for sources at the sky locations of interest.
We start by taking the expectation value of the optimal statistic,
\Deqn{Sopt}, using the optimal filter for a sky location $\hat \Omega$
assuming the redshift data contain a stochastic signal from that location. We
then divide by the square root of the variance
given in \Deqn{varianceSopt2}. The result is proportional to
\bea
G(\hat \Omega) = 
\left( \displaystyle\sum_{i=1}^{l}\displaystyle\sum_{j<i}^{l} 
{}^{(ij)}\Gamma_{\hat \Omega}^2\right)^{1/2}
\label{avsnr}
\eea 
where we have assumed (for illustrative purposes) that the noise
spectra of all pulsars is the same, the observation times for all
pairs is the same, and $n_{ij}=1$ for all pulsar pairs. 
Figure~\ref{fig:parkessens} shows the 
\Deqn{avsnr}, for the 20 pulsars of the Parkes Pulsar Timing
Array~\cite{Manchester:2007mx,PPTA}. Since most of the pulsars are in
the Southern hemisphere the Parkes Pulsar Timing
Array is most sensitive in that region.

\begin{figure}[t]
\centering
\subfigure{\includegraphics[width=3.7in]{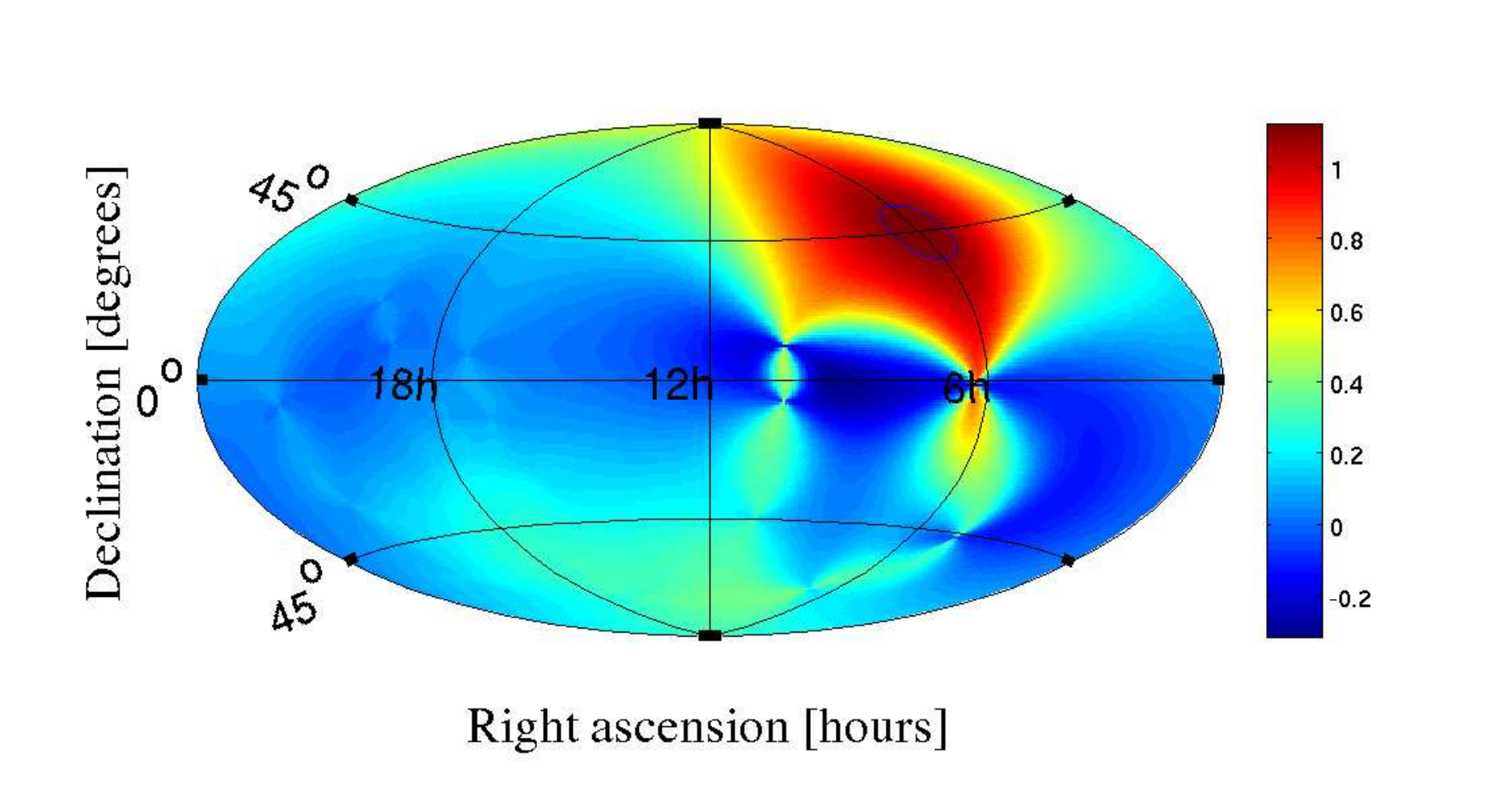}}\\
\subfigure{\includegraphics[width=3.7in]{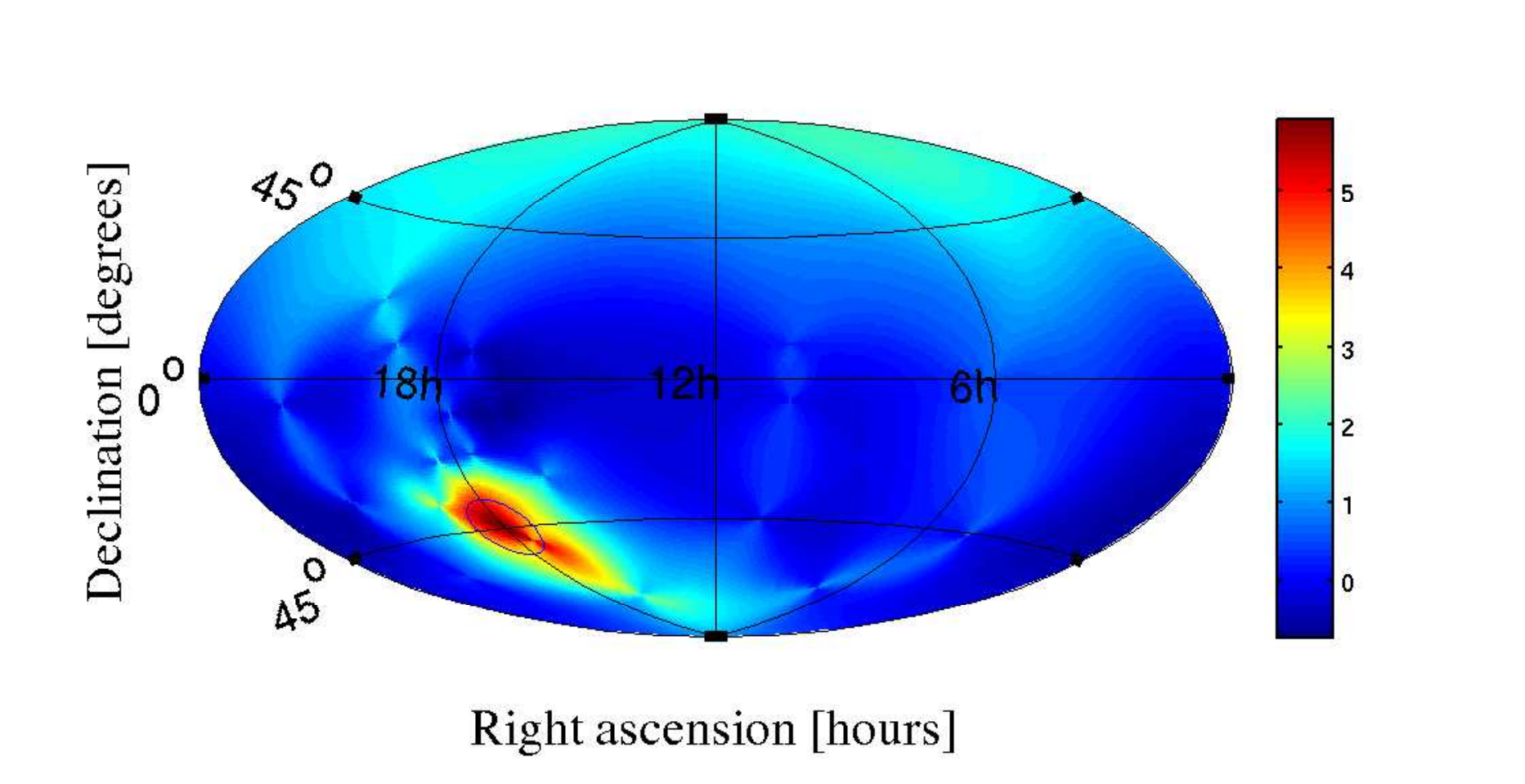}}
\caption{
Plot of the point spread function \Deqn{PSF} for the Parkes pulsar
timing array for a source at 6h RA 45$^\circ$ Dec (top panel) 
and 18h RA -45$^\circ$ Dec (bottom panel).}
\label{fig:parkesPSF}
\end{figure}

Another quantity of interest is the point spread function, which
measures the intrinsic spatial correlation of the skymap, or
equivalently, the ability of a pulsar network to locate a stochastic
source of gravitational waves. We construct the point spread function
by computing the signal to noise for a source at some sky location
that we search for using the optimal filter for some other location.
In particular, we take the expectation value of the optimal statistic,
\Deqn{Sopt}, using the optimal filter for a sky location $\hat \Omega$
assuming the redshift data contain a signal from another location
$\hat \Omega'$, then we divide by the square root of the variance
given in \Deqn{varianceSopt2}. The result is proportional to
\bea
A(\hat \Omega, \hat \Omega') = 
\frac{\displaystyle\sum_{i=1}^{l}\displaystyle\sum_{j<i}^{l} 
{}^{(ij)}\Gamma_{\hat \Omega} \, {}^{(ij)}\Gamma_{\hat \Omega'}}
{\left( \displaystyle\sum_{i=1}^{l}\displaystyle\sum_{j<i}^{l} 
{}^{(ij)}\Gamma_{\hat \Omega}^2\right)^{1/2}}
\label{PSF}
\eea
where we have again assumed that the noise
spectra of all pulsars is the same, the observation times for all
pairs is the same, and $n_{ij}=1$ for all pulsar pairs.
Figure~\ref{fig:parkesPSF} shows the point spread function,
\Deqn{PSF}, for the 20 pulsars of the Parkes Pulsar Timing
Array~\cite{Manchester:2007mx,PPTA} for a source at 6h RA 45$^\circ$
Dec (top panel) and another at 18h RA -45$^\circ$ Dec (bottom panel).

The point spread function can be understood in terms of the likelihood ratio
of Sec.~\ref{lhood}:  Suppose that the likelihood ratio is computed using
the overlap reduction function $\Gamma_{\hat\Omega}$ appropriate for a
stochastic signal coming from direction $\Gamma_{\hat\Omega}$ when the true
signal is in fact coming from direction $\Gamma_{\hat\Omega}$.  The the
expectation value of the log-likelihood ratio is [cf.~Eq.~(\ref{e:expectlikelihood})]
\begin{eqnarray}
  \langle\ln\Lambda\rangle 
  &=& \epsilon^2 \epsilon^2_{\mathrm{true}}
  \sum_{i=1}^l \sum_{j<i}^l {}^{(ij)}\Gamma_{\hat\Omega}\,{}^{(ij)}\Gamma_{\hat\Omega'}\,{}^{(ij)}C \nonumber\\
  && - {\textstyle\frac12}\epsilon^4
  \sum_{i=1}^l \sum_{j<i}^l {}^{(ij)}\Gamma_{\hat\Omega}^2\,{}^{(ij)}C
  \nonumber\\
  &&+ O(\epsilon^6)
\end{eqnarray}
with
\begin{equation}
  {}^{(ij)}C = \left(\frac{3H_0}{32\pi^3}\right)^2 T
  \int_{-\infty}^{\infty} df\,\frac{\hat\Omega_{\mathrm{gw}}^2(|f|)}{f^6P_i(|f|)P_j(|f|)}.
\end{equation}
If ${}^{(ij)}C$ is approximately the same for all pulsar pairs then
\begin{equation}
  \max_\epsilon \langle\ln\Lambda\rangle \propto A^2(\hat\Omega,\hat\Omega').
\end{equation}
In this sense the point spread function describes the degree to which the
position of a point source of stochastic gravitational waves can be located
in terms of the likelihood ratio.

\section{\label{issues} Issues with pulsar timing data}

We have derived the optimal statistic for detecting a stochastic background. 
In this section we would like to discuss some
issues associated with departures from the idealizations 
made to arrive at the optimal statistic.

\subsection{\label{CNsec}Colored noise and non-stationarity} 

In contrast to previous methods~\cite{Jenet:2005pv,Jenet:2006sv} the techniques 
presented here do not rely on the data being white.  The power spectra $P_i(|f|)$ 
in the optimal statistic account for colored noise. However, the methods assume the
data is stationary.  

If the data is non-stationary over long timescales
it can be divided into short stationary (or almost stationary) stretches and 
the power spectrum can be estimated using the Lomb-Scargle periodogram 
described below, or modeled for each stretch. The optimally filtered data 
stretches can then be 
combined along the lines discussed in Section~\ref{pta}. 
One concern associated with breaking the data up into small stretches is loss of 
low frequency information:  Gravitational waves with periods larger than the 
length of the short stretches will be lost in this procedure. 
The problem can be avoided by first computing the 
quantities $s_i(f)/P_i(|f|)$ for each of the short stretches and then combine them
using the Dirichlet kernel to  construct full time baseline versions of 
these quantities. 

If the spectrum is measured it can be smoothed by performing a
running average over a small frequency window, which if the data are stationary 
in the stretch the spectrum is estimated, is equivalent to ensemble averaging.

\subsection{\label{UnevenSec}Unevenly sampled data}

The fact that pulsar timing measurements are not taken continuously
leads to a data set that is unevenly sampled in time. This poses a
problem for frequency-domain analyses not present in their time-domain
counterparts. The authors of~\cite{Lommenea} will explore the time
domain approach in detail.  It is unclear which of these two methods
will turn out to be more robust.  In what follows we address the
specific issues of computing periodograms and Fourier transforms for
unevenly sampled data sets which we think is useful in any case.

\subsubsection{The Lomb-Scargle Periodogram}

The problem of constructing periodograms from unevenly sampled data comes up
in the data analysis of variable stars.  It was in precisely this context that
Lomb~\cite{Lomb:1976wy} and Scargle~\cite{Scargle:1982bw} proposed a 
least-squares solution to the problem.  
The basic idea is as follows:
Let $x(t_i)$ be a time series with zero mean sampled at $i=1\ldots N$ unevenly 
spaced times.  Now fit the time series by finding the coefficients $a_{\rm min}$ 
and $b_{\rm min}$ that minimize the square of the residual
\beq
r^2(f)\equiv\sum_{i=1}^N \{x(t_i) - a\cos[2\pi f(t_i-\tau)]
-b\sin[2\pi f(t_i-\tau)]\}^2,
\label{Rmin}
\eeq
where
\beq
\tan(4\pi f\tau)=
\frac{\sum_{i=1}^N\sin(4\pi ft_i)}{\sum_{i=1}^N\cos(4\pi ft_i)}.
\label{lstau}
\eeq
Then the periodogram is defined up to normalization by the difference
\bea
\Delta r^2(f) &=& \sum_{i=1}^N x^2(t_i) - r^2_{\rm min}(f)\nonumber \\
&=& \frac{\left(\sum_{i=1}^Nx(t_i)\cos\left[2\pi f(t_i-\tau)\right]\right)^2}{ 
\sum_{i=1}^N\cos^2\left[2\pi f(t_i-\tau)\right]}\nonumber\\ 
&\pheq&+
\frac{\left(\sum_{i=1}^Nx(t_i)\sin\left[2\pi f(t_i-\tau)\right]\right)^2}{ 
\sum_{i=1}^N\sin^2\left[2\pi f(t_i-\tau)\right]}
\eea
where $r^2_{\rm min}(f)$ is the quantity in \Deqn{Rmin} with $a=a_{\rm min}$ and
$b=b_{\rm min}$.  After normalization~\cite{Horne:1986bi} and generalization to 
data with nonzero mean, we have the Lomb-Scargle periodogram
\begin{widetext}
\beq
P_X^{\rm LS}(f) = \frac{1}{2\sigma_x^2}\left\{
\frac{\left(\sum_{i=1}^N[x(t_i)-\mu_x]\cos\left[2\pi f(t_i-\tau)\right]\right)^2}{
\sum_{i=1}^N\cos^2\left[2\pi f(t_i-\tau)\right]} 
+ \frac{\left(\sum_{i=1}^N[x(t_i)-\mu_x]\sin\left[2\pi f(t_i-\tau)\right]\right)^2}{
\sum_{i=1}^N\sin^2\left[2\pi f(t_i-\tau)\right]}\right\},
\eeq
\end{widetext}
where $\mu_x$ and $\sigma_x^2$ are the mean and variance, respectively of 
$x(t_i)$.  Note that the definition of $\tau$ in \Deqn{lstau} ensures that the
resulting periodogram is independent of where $t=0$. 

\subsubsection{Fourier transforms}

The idea of using a least-squares minimization is also useful for constructing 
Fourier transforms.  To do so, we borrow an idea from the radar 
community~\cite{HouseMount}.  Suppose  we have a timeseries, $x(t_i)$,
non-uniformly sampled at times $t_0\ldots t_N$ and we wish to construct its 
Fourier transform,
\beq
\tilde{x}(f_m) = \sum_{j=0}^N x(t_j) e^{-i2\pi f_mt_j}
\eeq
over $M$ evenly spaced frequencies, $f_0\ldots f_M$.  The strategy we will
employ is to use a least-squares procedure to find the best fit to the original
timeseries after an inverse Fourier transform.  That is, the (squared) residual 
to be minimized is given by
\beq
r^2=\sum_{j=0}^{N} \left|x(t_j) - \sum_{k=0}^M \tilde{\xi}(f_k) e^{i2\pi f_kt_j}\right|^2,
\label{ftres}
\eeq
where the $\tilde{\xi}(f_k)$ are to be determined.  Defining
\beq
A_{kj}=e^{i2\pi f_kt_j},
\eeq
we can write
\beq
r^2 = \left\lVert \vec{x} - A\vec{\tilde{\xi}}\right\rVert^2.
\eeq
The least-squares solution to this problem is given by
\beq
\vec{\tilde{\xi}} = \left(A^\dag A\right)^{-1}A^\dag\vec{x}.
\label{bft}
\eeq
The problem is then purely a computational one, which, because of the limited 
amount of pulsar timing data available, is completely tractable on a modern 
computer, regardless of the efficiency of the algorithm.  On a final note,
one can actually improve upon this procedure~\cite{HouseMount} by weighing the 
residual in \Deqn{ftres} by the square root of the 
variance, $\sigma_{x(t_j)}$, associated with each data point
\beq
r^2=\sum_{j=0}^{N} \frac{1}{\sigma_{x(t_j)}}
\left|x(t_j) - \sum_{k=0}^M \tilde{\xi}_k e^{i2\pi f_kt_j}\right|^2,
\eeq
which has the advantage of automatically including the
the error bars associated with individual pulsar timing data points.

\section{\label{SCsec}Summary and Conclusions}

A stochastic background of gravitational waves could be detected via
pulsar timing observations in the next 5 to 10 years.  This background
may be astrophysical, such as that produced by supermassive black
holes, or cosmological, such as that produced by a network of cosmic
(super)strings. In the latter case a detection would open a window
onto a time in the early universe prior to recombination and could
have profound consequences. Leveraging techniques developed for
ground-based instruments such as LIGO and Virgo, in this paper we have
shown how to optimally extract the signal produced by a stochastic
background of gravitational waves using cross-correlations of timing
data from a pulsar timing array.

We started by considering the redshift induced by a gravitational wave
on the frequency of arrival of radio pulses from a pulsar first derived
by Detweiler~\cite{Detweiler}. The redshift is proportional to the
difference in the metric perturbation at the pulsar (when a pulse is
emitted) and at the Earth (when that pulse is received).  Using a
convenient coordinate independent description of the signal we
examined the form of the signal in the frequency domain. The term
involving the metric perturbation at the pulsar is typically neglected
because it can be treated as a sort of noise term which averages to
zero in correlations of timing measurements of different pulsars.  In
the frequency domain the dependence on the metric perturbation at the
pulsar is in a phase factor that depends on the distance to the
pulsar. It is possible that if we could determine the distance to
pulsars with sufficient accuracy we could use the metric perturbation
at the pulsar to improve the sensitivity of our searches.
Unfortunately, accurate measurements of pulsar distances are
unavailable. By first finding the optimal cross-correlation filter, we
have shown that for pulsar distances and gravitational wave
frequencies typical of pulsar timing
experiments, the metric perturbation at the pulsar can be neglected
without a significant deviation from optimality. It is unclear whether
this is true for other types of gravitational wave searches.
We have also determined the optimal way to combine pulsar timing data
from a pulsar timing array, which is constructed from pairs of
optimally filtered cross-correlations.

We have discussed and illustrated frequentist and Bayesian methods for
setting upper limits using the distribution of the optimal statistic.
We have shown how to construct a pulsar timing radiometer: A map
of the sky created by optimizing the cross-correlation statistic for
particular sky directions. We have also shown how to determine the
intrinsic spatial correlation of such maps, which in turn determines
the ability of a pulsar timing array to locate a source of stochastic
gravitational waves.

We have ended with a discussion of some problems related to realistic
analysis of pulsar timing data, particularly the issues of
non-stationarity and uneven sampling. The optimal filter is
constructed from power spectra of the pulsar timing data, which can be
modeled or measured, and accounts for the effects of colored noise.
We have described a technique, the Lomb-Scargle periodogram, for
robust spectrum estimation that can be used to construct the optimal
filter. The optimal filter can then be applied in the frequency domain and
we have described a procedure for taking Fourier transforms of
unevenly sampled data that accounts for error bars in the individual
pulsar timing data points. The optimal filter can also be inverse
Fourier transformed and applied in the time domain where uneven
sampling is not an issue.  Regardless of which method turns out to be
more useful and robust for stochastic background searches, we believe
the development of Fourier techniques for unevenly sampled data will
be beneficial. Lommen, Romano and Woan~\cite{Lommenea} will examine
time-domain methods in detail.

\begin{acknowledgments} 
  We would like to thank Luis Anchordoqui, Steven Detweiler, Nick
  Fotopoulos, Rick Jenet, Adam Mercer, and Joe Romano for many useful
  discussions.  Additionally we would like to thank Nick Fotopoulos
  and Adam Mercer for carefully reading the manuscript. JC is
  supported in part by NSF Grant No. PHY-0701817. LP is supported by
  NSF Grant No. PHY-0503366 and the Research Growth Initiative at the
  University of Wisconsin- Milwaukee. XS is supported in part by NSF
  Grant No. PHY-0758155. SB gratefully acknowledges the support of the
  California Institute of Technology.  LIGO was constructed by the
  California Institute of Technology and Massachusetts Institute of
  Technology with funding from the National Science Foundation and
  operates under cooperative agreement PHY-0107417.
\end{acknowledgments}

\appendix
\section{\label{Detderiv} Derivation of Detweiler's formula using the 
geodesic equation}

For completeness of presentation, we include a derivation of Detweiler's
formula. We consider, as we did in Section~\ref{sigsec}, the metric perturbation
due to a single gravitational wave traveling in the $\hat{\Omega}=\hat{z}$ 
direction, so Eqs.~(\ref{mpzdir}-\ref{hpcdef}) hold. Then,  
if a vector $s^a$ is null in Minkowski space, the corresponding null vector,
$\sigma^a$, in the perturbed spacetime $g_{ab}=\eta_{ab}+h_{ab}$ is given 
by~\cite{Hellings83},
\begin{eqnarray}
\label{nullvectformula}
\sigma^a=s^a-\frac{1}{2}\eta^{ab}h_{bc}s^c.
\end{eqnarray}

The null vector in Minkowski space that points from the pulsar to the
solar system is $s^a=\nu (1,-\alpha,-\beta,-\gamma)$, where, as before, 
$\alpha$, $\beta$ and $\gamma$ are the direction cosines with the $x$-, $y$- and 
$z$-directions, respectively. The corresponding
perturbed vector is readily computed from \Deqn{nullvectformula}
\begin{equation}
\label{nullvector}
\sigma^a=\nu\left( \begin{array}{c}
1\\
-\alpha(1-\frac{1}{2}h_+)+\frac{1}{2}\beta  h_\times \\
-\beta (1+\frac{1}{2}h_+)+\frac{1}{2}\alpha h_\times \\
-\gamma\\
\end{array}\right).
\end{equation}

The geodesic equation tells us that the $t$-component of $\sigma^a$ satisfies
\begin{equation}
\label{nullvectgeodesiceqn}
\frac{d\sigma^t}{d\lambda}=-\Gamma^t_{ab}\sigma^a \sigma^b.
\end{equation}
It follows from the form of the metric perturbation in \Deqn{gmunu} that
\bea
\label{christoffel3}
\Gamma^t_{ab}&=&-\frac{1}{2}g^{tc}\left[
\frac{\partial g_{bc}}{\partial x^a} + \frac{\partial
  g_{a c}}{\partial x^b} -
\frac{\partial g_{ab}}{\partial x^c} \right]
\nonumber\\
&=&\frac{1}{2}\dot g_{ab}\nonumber\\
&=&\frac{1}{2}
\left(
\begin{array}{cccc}
0 & 0 & 0 & 0  \\
0 & \dot h_+ & \dot h_{\times} & 0 \\
0 & \dot h_{\times} & -\dot h_+ & 0 \\
0 & 0 & 0 & 0\\
\end{array}
\right),
\eea
where the overdot denotes a derivative with respect to $t$.
The geodesic equation then reads
\bea
\label{nullvectgeodesiceqn2}
\frac{d\sigma^t}{d\lambda}&=&-\frac{1}{2}\dot
g_{ab}\sigma^a \sigma^b\nonumber\\
&=&-\frac{1}{2}\left[ \dot g_{xx}
  (\sigma^x)^2 - \dot g_{yy} (\sigma^y)^2 \right] - \dot
g_{xy}\sigma^x \sigma^y\nonumber\\
&=&-\frac{1}{2}\dot h_+\left[ 
  (\sigma^x)^2 - (\sigma^y)^2 \right] + \dot h_\times \sigma^x \sigma^y.
\eea
After a little algebra \Deqn{nullvector} leads to
\begin{equation}
\label{orderh1}
(\sigma^x)^2 - (\sigma^y)^2 = \nu^2 \left( \alpha^2 - \beta^2 \right)
+ {\cal O}(h),
\end{equation}
as well as
\begin{equation}
\label{orderh2}
\sigma^x \sigma^y = \nu^2 \alpha\beta  + {\cal O}(h),
\end{equation}
so that
\begin{equation}
\label{nullvectgeodesiceqn4}
-\frac{d\nu}{d\lambda}=\frac{1}{2}\dot h_+ \nu^2 \left( 
  \alpha^2 - \beta^2 \right) + \dot h_\times \nu^2 \alpha \beta.
\end{equation}
We proceed by writing the time derivatives in
\Deqn{nullvectgeodesiceqn4} as derivatives with respect to the
affine parameter $\lambda$. In particular, since
$h_{+,\times}=h_{+,\times}(t-z)$ we have
\begin{equation}
\label{partialderiv1}
\frac{d h_{+,\times}}{d\lambda}=\frac{\partial h_{+,\times}}{\partial
  t} \frac{d t}{d\lambda}+\frac{\partial h_{+,\times}}{\partial
  z} \frac{d z}{d\lambda}.
\end{equation} 
Moreover, we also have that the frequency $\nu=dt/d\lambda$, in addition to 
$\partial h_{+,\times}/ \partial z = - \partial h_{+,\times}/ \partial t$, and
$dz/d\lambda=-\nu \gamma$. Therefore we can write \Deqn{partialderiv1} as
\begin{equation}
\label{partialderiv3}
\dot h_{+,\times} = \frac{1}{\nu (1+\gamma)} \frac{d h_{+,\times}}{d\lambda}.
\end{equation}
Then \Deqn{nullvectgeodesiceqn4}, the geodesic equation, becomes 
\begin{equation}
\label{redshift1}
-\frac{1}{\nu}\frac{d\nu}{d\lambda}=\frac{1}{2} \frac{  
  \alpha^2 - \beta^2 }{1+\gamma} \frac{d h_{+}}{d\lambda}+
\frac{\alpha \beta}{1+\gamma}
\frac{d h_{\times}}{d\lambda},
\end{equation}
which we integrate to find
\begin{equation}
\label{redshift2}
\frac{\nu(t)}{\nu_0} = \exp \left[ -\frac{1}{2} \frac{  
  \alpha^2 - \beta^2 }{1+\gamma} \Delta h_{+}-
\frac{\alpha \beta}{1+\gamma}
 \Delta h_{\times} \right].
\end{equation}
It is worth pointing out that the direction cosines are functions of
the affine parameter. The dependence is in terms of ${\cal O}(h)$, and
we have neglected this dependence in going from \Deqn{redshift1} to
\Deqn{redshift2}. The final result is obtained by expanding this
expression to first order in $h_{+,\times}$
\begin{equation}
\label{detweilerresult}
\frac{\nu_0-\nu(t)}{\nu_0} =  \frac{1}{2} \frac{  
  \alpha^2 - \beta^2 }{1+\gamma} \Delta h_{+}+
\frac{\alpha \beta}{1+\gamma}
 \Delta h_{\times},
\end{equation} 
where $\Delta h_{+,\times}=h^{\rm p}_{+,\times}-h^{\rm e}_{+,\times}$ is the difference 
between the metric perturbation at the pulsar and the detector. 
This expression is precisely \Deqn{redshift}.

\section{Linearity of the redshift}
\label{app:lin}
In this appendix we will explicitly demonstrate that the total
redshift is the sum of the contributions from gravitational waves in
every direction, as written in \Deqn{zftot}.  The derivation is merely
a generalization of the results derived in the previous appendix.  We begin by
considering a metric perturbation, $h_{ab}$, in a spatial
transverse-traceless gauge comprised of the sum of metric
perturbations, $h_{ab}^{(i)}$, from gravitational waves in $N$
different directions, $\hat{\Omega}_{(i)}$. Namely,
\beq
h_{ab} = \sum_{(i)}^{N}h_{ab}^{(i)}(t-\hat{\Omega}_{(i)}\cdot\vec{x}),
\label{mpsum}
\eeq  
where $t$ and $\vec{x}$ form a four vector, $x^a$, in the background (Minkowski)
geometry. Adjusting the notation from Appendix A,
\bea
 s^a &=& dx^a/d\lambda\nonumber\\ 
&=& \nu(1,-\alpha,-\beta,-\gamma)\nonumber\\
&\equiv& \nu(1,-\hat{p}),
\label{sdef}
\eea
as a null vector in Minkowski space.  The null geodesic is perturbed by \Deqn{mpsum}, resulting in a
\beq
\sigma^a = s^a +\delta s^a.
\eeq
As before, our interest is in the quantity
\beq
\frac{d\sigma^t}{d\lambda}=-\Gamma^t_{ab}\sigma^a \sigma^b.
\eeq
The spatial nature of the gauge we've chosen ensures that
\bea
\Gamma^t_{ab} &=& \frac{1}{2}\dot{g}_{ab}\\
&=&\frac{1}{2}\dot{h}_{ab},
\eea
which is evident from the first line of \Deqn{christoffel3}.  It follows from these
definitions that
\bea
\frac{d\sigma^t}{d\lambda}&=&-\frac{1}{2}\dot{h}_{ab}\sigma^a \sigma^b
\nonumber\\
&=&-\frac{1}{2}\dot{h}_{ab}(s^a+\delta s^a)(s^b+\delta s^b)
\nonumber\\
&=&-\frac{1}{2}\dot{h}_{ab}s^as^b
\nonumber\\
&=&-\frac{1}{2}\dot{h}_{ij}(\nu^2 p^ip^j),
\label{dsigmagen}
\eea
where $i$ and $j$ are spatial indices.  As before, we want to write the expression 
above in terms of the affine parameter along $s^a$.  We begin by noting that for term in
\Deqn{mpsum}
\bea
\frac{d h_{ab}^{(i)}(t-\hat{\Omega}_{(i)}\cdot\vec{x})}{d\lambda} &=& 
\frac{\partial h_{ab}^{(i)}(t-\hat{\Omega}_{(i)}\cdot\vec{x})}{\partial t} \frac{dt}{d\lambda}\nonumber\\
&\pheq&
+ \frac{\partial h_{ab}^{(i)}(t-\hat{\Omega}_{(i)}\cdot\vec{x})}{\partial (\hat{\Omega}_{(i)}\cdot\vec{x})} 
\frac{d(\hat{\Omega}_{(i)}\cdot\vec{x})}{d\lambda}\nonumber\\
&=& \frac{\partial h_{ab}^{(i)}(t-\hat{\Omega}_{(i)}\cdot\vec{x})}{\partial t}\nu\nonumber\\
&\pheq&
- \frac{\partial h_{ab}^{(i)}(t-\hat{\Omega}_{(i)}\cdot\vec{x})}{\partial t}
\hat{\Omega}_{(i)}\cdot\frac{d\vec{x}}{d\lambda}\nonumber\\
&=& \nu(1+\hat{\Omega}_{(i)}\cdot\hat{p})\frac{\partial h_{ab}^{(i)}(t-\hat{\Omega}_{(i)}\cdot\vec{x})}{\partial t},
\nonumber\\
\eea
where we have used the $(t-\hat{\Omega}_{(i)}\cdot\vec{x})$ dependence of the metric perturbation to write the
spatial derivatives as time derivatives along with \Deqn{sdef}.  Putting this together with \Deqn{dsigmagen}, 
the result is that
\beq
-\frac{1}{\nu}\frac{d\nu}{d\lambda} = \sum_{(i)}^N\frac{1}{2}\frac{\hat{p}^i\hat{p}^j}{1+\hat{\Omega}_{(i)}\cdot\hat{p}} 
h_{ij}^{(i)}(t-\hat{\Omega}_{(i)}\cdot\vec{x}),
\eeq
which can be integrated to give the redshift
\beq
z \equiv \frac{\nu_0-\nu(t)}{\nu_0} = \sum_{(i)}^N\frac{1}{2}\frac{\hat{p}^i\hat{p}^j}{1+\hat{\Omega}_{(i)}\cdot\hat{p}} 
\Delta h_{ij}^{(i)}(t-\hat{\Omega}_{(i)}\cdot\vec{x}),
\eeq
which is the discrete version of \Deqn{zftot}.

\section{The Overlap Reduction Function in the High-Frequency Limit}
\subsection{Derivation of the Hellings-Downs curve}
\label{app:HDC}
In this section we derive the Hellings and Downs curve given by Eq.~(\ref{orfapprox})~\cite{HD}.  We begin with the definition of the overlap reduction
function, Eq.~(\ref{orf}), and we ignore the exponential factors.  Thus we
wish to evaluate
\begin{equation}
  \Gamma_0 = \beta\sum_{A=+,\times}\int_{S^2}d\hat\Omega\,
  F_1^A(\hat\Omega) F_2^A(\hat\Omega)
\end{equation}
and using the definition of $F^A(\hat\Omega)$ given by Eq.~(\ref{eqFA}) we find
\begin{equation}\label{indexORF}
  \Gamma_0 = \frac{1}{4}\beta\sum_{A=+,\times}\int_{S^2}d\hat\Omega\,
  \frac{\hat{p}_1^i\hat{p}_1^j}{1+\hat\Omega\cdot\hat{p}_1}
  \frac{\hat{p}_2^k\hat{p}_2^l}{1+\hat\Omega\cdot\hat{p}_2}
  e^A_{ij}(\hat\Omega)e^A_{kl}(\hat\Omega).
\end{equation}
The two unit vectors $\hat{p}_1$ and $\hat{p}_2$ are those pointing from the
Earth toward the first and second pulsar respectively and the polarization
tensors $e^+_{ij}(\hat\Omega)$ and $e^\times_{ij}(\hat\Omega)$ for a
gravitational wave traveling in direction $\hat\Omega$ are given by
Eqs.~(\ref{eq:e_plus}) and~(\ref{eq:e_cross}) respectively.  To evaluate the
integral we choose a coordinate system in which $\hat{p}_1$ is parallel to the
$z$-axis and $\hat{p}_2$ is in the $x$-$z$ plane.  Then
\begin{subequations}
\begin{align}
\label{eq:p1}
  {\hat{p}}_1&= (0,  0,  1) \\
\label{eq:p2}
  {\hat{p}}_2&=(\sin{\xi}, 0, \cos{\xi})
\end{align}
\end{subequations}
where $\xi$ is the angular separation between the two pulsars.  Because
we have chosen coordinates in which $\hat{p}\cdot\hat{m}=0$
[cf. Eq.~(\ref{eq:m})], the $\times$-polarization terms vanish and our
expression for $\Gamma_0$ becomes
\begin{widetext}
\begin{equation}
  \Gamma_0 = -\frac{1}{4}\beta \int_{S^2} d{\hat\Omega}\,\frac{\sin^2\theta\left(\sin^2\xi\sin^2\phi - \sin^2\xi\cos^2\theta\cos^2\phi - \cos^2\xi\sin^2\theta + 2 \sin\xi \cos\xi \sin\theta \cos\theta \cos\phi\right) }
{(1+ \cos\theta)(1+\cos\xi\cos\theta + \sin\xi\sin\theta\cos\phi)}
\end{equation}
\end{widetext}
Straightforward manipulation shows that this integral becomes
\begin{equation}\label{e:orf0IJ}
  \Gamma_0 = \frac{1}{4}\beta(I+J)
\end{equation}
with
\begin{equation}\label{e:I}
\begin{split}
  I &= \int_{S^2} d\hat\Omega\,(1-\cos\theta)
  (1-\cos\xi\cos\theta - \sin\xi\sin\theta\cos\phi)\\
  &= 4\pi\left(1+\frac{1}{3}\cos\xi\right)
\end{split}
\end{equation}
and
\begin{equation}\label{Jdef}
  J = -2\sin^2\xi\int_0^\pi d\theta\,\sin\theta(1-\cos\theta) K \\
\end{equation}
where we have defined
\begin{equation}\label{Keq}
K \equiv \int_0^{2\pi}d\phi\,
  \frac{\sin^2\phi}{1+\cos\xi\cos\theta+\sin\xi\sin\theta\cos\phi}.
\end{equation}
K may be trivially evaluated by contour integration in the complex plane.  The result is
\begin{equation}\label{Kresult}
\begin{split}
K &= 2\pi\frac{1+\cos\xi\cos\theta + |\cos\xi+\cos\theta|}{\sin^2\xi\sin^2\theta}\\
  &= 2\pi\left(\frac{1\mp\cos\xi}{\sin^2\xi}\right) \left(\frac{1\mp\cos\theta}{\sin^2\theta}\right)
\end{split}
\end{equation}
where the negative sign applies when $0<\theta<\pi-\xi$ and the positive
sign applies when $\pi-\xi<\theta<\pi$.
Hence we find that
\begin{equation}\label{e:J}
\begin{split}
  J =& -4\pi(1-\cos\xi)\int_0^{\pi-\xi}d\theta\,
  \frac{(1-\cos\theta)^2}{\sin\theta} \\
  &-4\pi(1+\cos\xi)\int_{\pi-\xi}^\pi d\theta\,\sin\theta \\
  =&\, 16\pi(1-\cos\xi)\ln\left(\sin\frac{\xi}{2}\right).
\end{split}
\end{equation}
Combining Eqs.~(\ref{e:orf0IJ}), (\ref{e:I}), and~(\ref{e:J}), we obtain
\begin{equation}
\begin{split}
  \Gamma_0 &= \frac{4\pi}{3}\beta\left\{1 + 3(1-\cos\xi)\left[
  \ln\left(\sin\frac{\xi}{2}\right) - \frac{1}{12}\right]\right\} \\
  &= \frac{4\pi}{3}\beta\left\{1 + \frac{3}{2}(1-\cos\xi)\left[
  \ln\left(\frac{1-\cos\xi}{2}\right) - \frac{1}{6}\right]\right\}.
\end{split}
\end{equation}
The expression in braces achieves a maximum value of unity when $\xi=0$, so
the correct normalization constant is $\beta=3/4\pi$.  With this normalization
we recover Eq.~(\ref{orfapprox}).
\subsection{Generalization to a Dipole Stochastic Background}
\label{app:dipole}
We now generalize the Hellings-Downs curve to the case of a stochastic background with a dipole moment in the direction $\hat{D}$.  We will start by defining the following quantities:
\begin{equation}
\hat{D} = \left(\sin\alpha_1\cos\eta, \sin\alpha_1\sin\eta, \cos\alpha_1\right)
\end{equation}
\begin{equation}
\begin{split}
  \hat{D}\cdot\hat{\Omega} &\equiv \cos\chi \\
  &= \cos\alpha_1\cos\theta + \sin\alpha_1\sin\theta\cos(\phi - \eta)
\end{split}
\end{equation}
\begin{subequations}
\begin{align}
\hat{D}\cdot\hat{p_1} &\equiv \cos\alpha_1\\
\begin{split}
\hat{D}\cdot\hat{p_2} &\equiv \cos\alpha_2 \\
&=  \cos\alpha_1\cos\xi + \sin\alpha_1\sin\xi\cos\eta
\end{split}
\end{align}
\end{subequations}
This derivation differs from the derivation of the Hellings-Downs curve only in that a factor $\hat{D}\cdot\hat\Omega$ must be included in the integral.  
\begin{equation}
  \Gamma_{\rm dip} = \frac{1}{4}\beta\sum_{A=+,\times}\int_{S^2}d\hat\Omega\,
  \frac{\hat{p}_1^i\hat{p}_1^j\,
    \hat{p}_2^k\hat{p}_2^l e^A_{ij}(\hat\Omega)e^A_{kl}(\hat\Omega)}{(1+\hat\Omega\cdot\hat{p}_1)(1+\hat\Omega\cdot\hat{p}_2)}\hat{D}\cdot\hat{\Omega}.
\end{equation}
This integral can be written as
\begin{widetext}
\begin{equation}\label{hunormousintegral}
\begin{split}
  \Gamma_{\rm dip} =& -\frac{1}{4}\beta \int_{S^2} d{\hat\Omega}\,\cos\chi \frac{\sin^2\theta\left(\sin^2\xi\sin^2\phi - \sin^2\xi\cos^2\theta\cos^2\phi - \cos^2\xi\sin^2\theta + 2 \sin\xi \cos\xi \sin\theta \cos\theta \cos\phi\right) }
{(1+ \cos\theta)(1+\cos\xi\cos\theta + \sin\xi\sin\theta\cos\phi)}.
\end{split}
\end{equation}
As in the previous section, we write
\begin{equation}
  \Gamma_{\rm dip} = \frac{1}{4}\beta(I+J)
\end{equation}
where the first term is now given by
\begin{equation}
\begin{split} I &= \int_{S^2} d\hat\Omega\,\cos\chi(1-\cos\theta)
  (1-\cos\xi\cos\theta - \sin\xi\sin\theta\cos\phi)\\
  &= -\frac{4\pi}{3}\left(\cos\alpha_1 + \cos\alpha_2\right)
\end{split}
\end{equation}
and $J$ is as in Eq.~(\ref{Jdef}), but $K$ is now given by
\begin{equation}
K = \int_0^{2\pi}d\phi\,
  \frac{\sin\xi\cos\alpha_1\cos\theta\sin^2\phi + (\cos\alpha_2 - \cos\alpha_1\cos\xi)\sin\theta \left(\cos\phi - \sin\phi\right)\sin^2\phi}{\sin\xi(1+\cos\xi\cos\theta)+ \sin^2\xi\sin\theta\cos\phi}
\end{equation}
and may be evaluated by the same methods.  The result is
\begin{equation}
K = \frac{2\pi}{\sin^2\xi}\left(a_{\pm}\cot\theta\csc\theta + b_{\pm}\cot^2\theta + c_{\pm}\csc^2\theta\right)
\end{equation}
where the following constant terms have been defined:
\begin{subequations}
\begin{align}
a_{\pm} &= \cos\alpha_1(1 \mp \cos\xi) \pm \frac{\cos\alpha_2 - \cos\alpha_1\cos\xi}{\sin^2\xi}\left(1 \mp \cos\xi\right)^2\\
b_{\pm} &= \mp \cos\alpha_1(1 \mp \cos\xi) - \frac{\cos\alpha_2 - \cos\alpha_1\cos\xi}{2 \sin^2\xi}\left(1 \mp \cos\xi\right)^2\\
c_{\pm} &=  - \frac{\cos\alpha_2 - \cos\alpha_1\cos\xi}{2 \sin^2\xi}\left(1 \mp \cos\xi\right)^2
\end{align}
\end{subequations}
and $a_{+}$, $b_{+}$, and $c_{+}$ are to be used in the case where the inequality $ 0 < \theta < \pi - \xi $ holds, and $a_{-}$, $b_{-}$, and $c_{-}$ are to be used otherwise.  Thus, the integral J must again be split into two sections, and the result of the integration is
\begin{equation}
J = 4\pi \left(\cos\xi - 1\right)\left(\cos\alpha_1 + \cos\alpha_2\right) - 16\pi \left(\cos\alpha_1 + \cos\alpha_2\right)\tan^2\frac{\xi}{2} \ln\left(\sin\frac{\xi}{2}\right)
\end{equation}
Thus, we see that 
\begin{equation}
  \Gamma_{\rm dip} = \pi\beta\left(\cos\alpha_1 + \cos\alpha_2\right)\left(\cos\xi - \frac{4}{3} - 4\tan^2\frac{\xi}{2}  \ln\left(\sin\frac{\xi}{2}\right)\right)
\end{equation}
Because we wish for $\Gamma_{\rm dip}$ to have maximal value of unity at $\xi = 0$ and $\xi = \pi$ (where it is clear that $\alpha_1 = \alpha_2 = 0$), we must select a normalization constant of $\beta = -3 / 2\pi$.

\end{widetext}

\end{document}